\documentclass[twocolumn]{article}

\usepackage{setspace}

\usepackage{amssymb}

\usepackage{graphicx}
\usepackage{amsmath}
\usepackage{float}

\begin{document}

\singlespacing

%\doublespacing

\onecolumn
%\begin{frontmatter}
\title{Recent advances in the internal functionalization of carbon nanotubes: synthesis, optical, and magnetic resonance studies}
\author{Ferenc Simon \dag, Rudolf Pfeiffer, and Hans Kuzmany \\ \textit{Institut f\"{u}r Materialphysik, Universit\"{a}t
Wien }\\ \textit{Strudlhofgasse 4, A-1090 Wien, Austria} \\
\texttt{ferenc.simon@univie.ac.at} \\
%\bigskip\\ Date of receiving: 28 March 2006, date of acceptance: 13 June 2006
}

%\affiliation{Institut f\"{u}r Materialphysik, Universit\"{a}t
%Wien, Strudlhofgasse 4, A-1090 Wien, Austria}

\date{}

%\newpage
\maketitle

%\newpage

%\bigskip
%\textit{Key words} Carbon nanotubes  Fullerene encapsulation
%Double-wall carbon nanotubes  Isotope engineered nanotubes Raman
%spectroscopy  Magnetic resonance spectroscopy

%\end{frontmatter}

%\newpage

%\tableofcontents

\begin{abstract}
The hollow inside of single-wall carbon nanotubes (SWCNT) provides a
unique degree of freedom to investigate chemical reactions inside
this confined environment and to study the tube properties. It is
reviewed herein, how encapsulating fullerenes, magnetic fullerenes,
$^{13}$C isotope enriched fullerenes and organic solvents inside
SWCNTs enables to yield unprecedented insight into their electronic,
optical, and interfacial properties and to study their growth.
Encapsulated C$_{60}$ fullerenes are transformed to inner tubes by a
high temperature annealing. The unique, low defect concentration of
inner tubes makes them ideal to study the effect of diameter
dependent treatments such as opening and closing of the tubes. The
growth of inner tubes is achieved from $^{13}$C enriched
encapsulated organic solvents, which shows that fullerenes do not
have a distinguished role and it opens new perspectives to explore
the in-the-tube chemistry. Encapsulation of magnetic fullerenes,
such as N@C$_{60}$ and C$_{59}$N is demonstrated using ESR. Growth
of inner tubes from $^{13}$C enriched fullerenes provides a unique
isotope engineered heteronuclear system, where the outer tubes
contain natural carbon and the inner walls are controllably $^{13}$C
isotope enriched. The material enables to identify the vibrational
modes of inner tubes which otherwise strongly overlap with the outer
tube modes. The $^{13}$C NMR signal of the material is specific for
the small diameter SWCNTs. Temperature and field dependent $^{13}$C
$T_1$ studies show a uniform metallic-like electronic state for all
inner tubes and a low energy, ~3 meV gap is observed that is
assigned to a long sought Peierls transition.
\end{abstract}

\tableofcontents

\newpage

% \maketitle

\twocolumn

\section{Introduction}

The era of nanotechnology received an enormous boost with the
discovery of carbon nanotubes (CNTs) by Sumio Iijima in 1991
\cite{IijimaNAT1991}. Before 1991 nano- and nanotechnology usually
meant small clusters of atoms or molecules. The originally
discovered CNTs contain several coaxial carbon shells and are called
multi-wall CNTs (MWCNTs). Soon thereafter single-wall CNTs (SWCNTs),
i.e. a carbon nanotube consisting of a single carbon shell were
discovered \cite{IijimaNAT1993,BethuneNAT1993}. The principal
interest in CNTs comes from the fact that they contain carbon only
and all carbon are locally sp$^{2}$ bound, like in graphite, which
provides unique mechanical and transport properties. This, combined
with their huge, $>$ 1000, aspect ratio (the diameters being 1-20 nm
and their lengths over a few micron or even exceeding cms) gives
them an enormous application potential. The not exhaustive list of
applications includes field-emission displays (epxloiting their
sharp tips) \cite{Obraztsov}, cathode emitters for small sized x-ray
tubes for medical applications \cite{ZhouAPL2002}, reinforcing
elements for CNT-metal composites, tips for scanning probe
microscopy \cite{HafnerNAT}, high current transmitting wires, cables
for a future space elevator, elements of nano-transistors
\cite{BachtoldSCI2001}, and elements for quantum information
processing \cite{HarneitPSS}.

Carbon nanotubes can be represented as rolled up graphene sheets,
i.e. single layers of graphite. Depending on the number of coaxial
carbon nanotubes, they are usually classified into multi-wall carbon
nanotubes (MWCNTs) and single-wall carbon nanotubes (SWCNTs). Some
general considerations have been clarified in the past 14 years of
nanomaterial research related to these structures. MWCNTs are more
homogeneous in their physical properties as the large number of
coaxial tubes smears out individual tube properties. This makes them
suitable candidates for applications where their nanometer size and
the conducting properties can be exploited such as e.g. nanometer
sized wires. In contrast, SWCNT materials are grown as an ensemble
of weakly interacting tubes with different diameters. The physical
properties of similar diameter SWCNTs can change dramatically as the
electronic structure is very sensitive on the rolling-up direction,
the so-called chiral vector \cite{HamadaPRL1992,DresselhausTubes}.
The chiral vector is characterized by the $(n,m)$ vector components
which denote the direction along which a graphene sheet is rolled up
to form a nanotube. Depending on the chiral vector, SWCNTs can be
metallic or semiconducting \cite{DresselhausTubes}. This provides a
richer range of physical phenomena as compared to the MWCNTs,
however significantly limits the range of applications. To date,
neither the directed growth nor the controlled selection of SWCNTs
with a well defined chiral vector has been performed successfully.
Thus, their broad applicability is still awaiting. Correspondingly,
current research is focused on the post-synthesis
separation of SWCNTs with a narrow range of chiralities \cite%
{ChattopadhyayJACS,KrupkeSCI,RinzlerNL,StranoSCI} or on methods
which yield information that are specific to SWCNTs with different
chiralities. Examples for the latter are the observation of
chirality selective band-gap fluorescence in semiconducting SWCNTs
\cite{Bachilo:Science298:2361:(2002)} and chirality assigned
resonant Raman scattering \cite{FantiniPRL2004,TelgPRL2004}.

Clearly, several fundamental questions need to be answered before
all the benefits of these novel nanostructures can be fully
exploited. Recent theoretical and experimental efforts focused on
the understanding of the electronic and optical properties of
single-wall carbon nanotubes. It has been long thought that the
one-dimensional structure of SWCNTs renders their electronic
properties inherently one-dimensional
\cite{HamadaPRL1992,DresselhausTubes}. This was suggested to result
in a range of exotic correlated phenomena such as the
Tomonaga-Luttinger (TLL) state \cite{EggerPRL1997}, the Peierls
transition \cite{BohnenPeierlsPRL2004,ConnetablePeierlsPRL2005},
ballistic transport \cite{DekkerNAT1997}, and bound excitons
\cite{KanePRL2003,LouiePRL2004,AvourisPRL2004,AvourisPRL2005}. The
presence of the TLL state is now firmly established
\cite{BockrathNAT,KatauraNAT2003,PichlerPRL2004}, there is evidence
for the ballistic transport properties \cite{DekkerNAT1997} and
there is growing experimental evidence for the presence of excitonic
effects \cite{HeinzSCI2005,MaultzschPRB2005}. The Peierls
transition, however remains still to be seen.

\begin{figure}
\begin{center}
\includegraphics[width=0.9\linewidth]{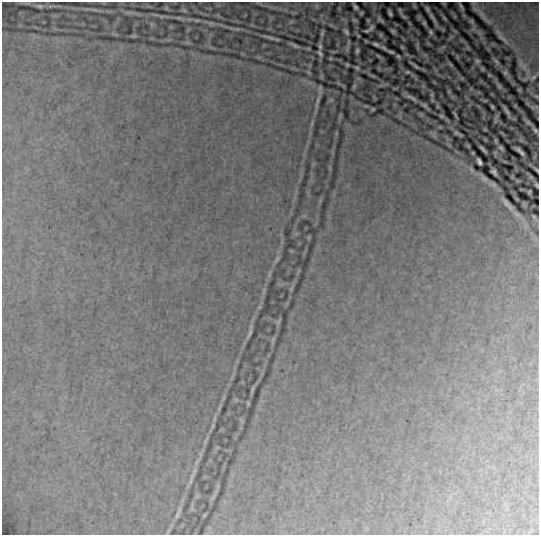}
\caption{HR-TEM image of C$_{60}$@SWCNT peapods.}
\label{CACReview_PeapodHRTEM}
\end{center}
\end{figure}

An appealing tool to study the SWCNT properties originates from the
discovery of fullerenes encapsulated inside SWCNTs by Smith,
Monthioux, and Luzzi \cite{SmithNAT}. This peapod structure is
particularly interesting as it combines two fundamental forms of
carbon: fullerenes and carbon nanotubes. A high-resolution
transmission electron microscopy (HR-TEM) image of a peapod is shown
in Fig. \ref{CACReview_PeapodHRTEM}. It was also shown that
macroscopic filling with the fullerenes can be achieved
\cite{LuzziCPL1999,KatauraSM2001}. This, in principle, opens the way
to encapsulate magnetic fullerenes which would enable the study of
the tube electronic properties using electron spin resonance as it
is discussed in this review. Another interesting follow-up of the
peapod structure discovery is that the encapsulated fullerenes can
be fused into a smaller diameter inner tube
\cite{LuzziCPL2000,BandowCPL2001} thus producing a double-wall
carbon nanotube (DWCNT). DWCNTs were first observed to form under
intensive electron radiation \cite{LuzziCPL1999} in a high
resolution transmission electron microscope from C$_{60}$ peapods.
Following the synthesis of C$_{60}$ peapods in macroscopic amounts
\cite{KatauraSM2001}, bulk quantities of the DWCNT material are
available using a high temperature annealing method
\cite{BandowCPL2001}. Alternatively, DWCNTs can be produced with
usual synthesis methods such as arc-discharge
\cite{HutchisonCAR2001} or CVD \cite{ChengCPL2002} under special
conditions. According to the number of shells, DWCNTs are between
SWCNTs and MWCNTs. Thus, one expects that DWCNTs may provide a
material where improved mechanical stability as compared to SWCNTs
coexists with the rich variety of electronic properties of SWCNTs.
There are, of course, a number of yet unanswered questions e.g. if
the outer tube properties are unaffected by the presence of the
inner tube or if the commensurability of the tube structures plays a
role. These questions should be answered before the successful
application of these materials.

\begin{figure}
\begin{center}
\includegraphics[width=0.9\linewidth]{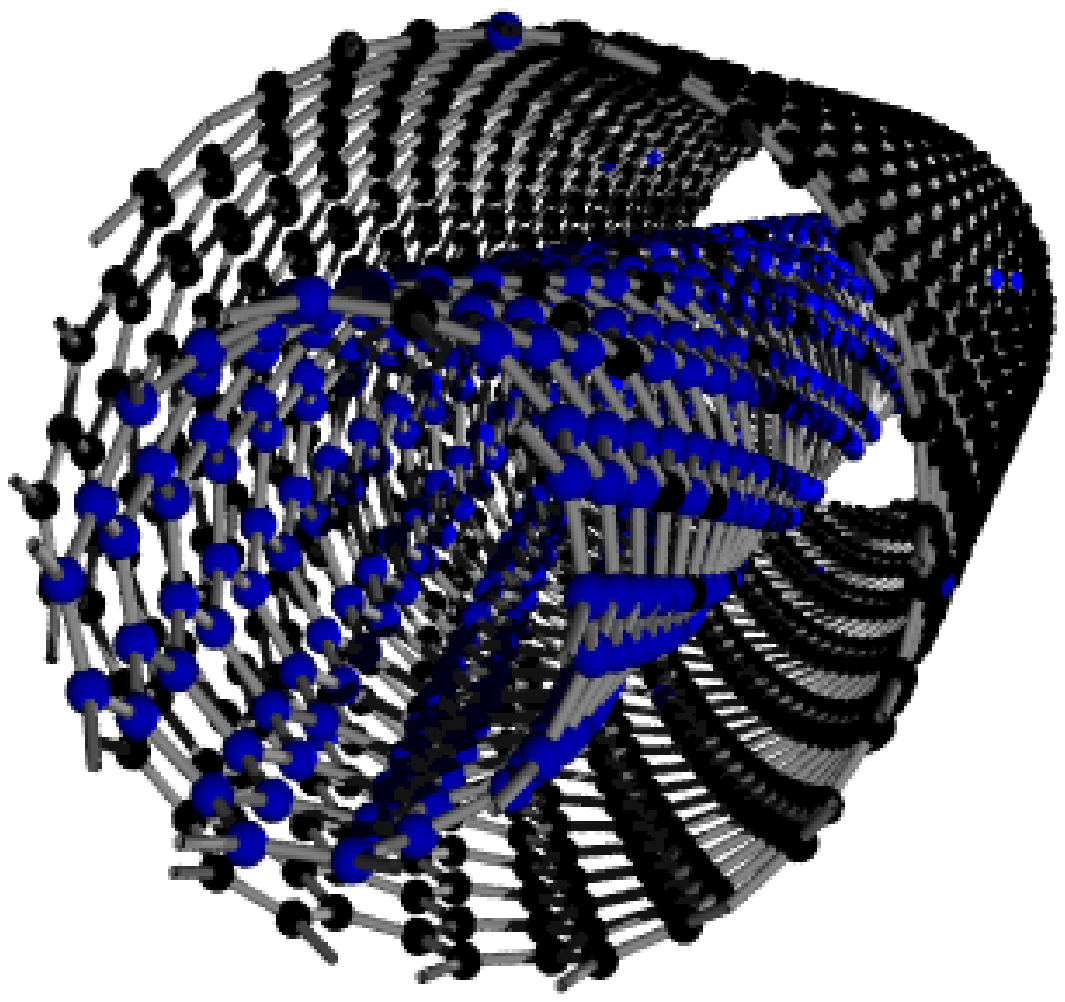}
\caption{Schematic structure of an isotope engineered DWCNT with
(14,6) outer and (6,4) inner tubes. $^{12}$C and $^{13}$C are shown
in black and blue, respectively. The inner tube is 89 \% $^{13}$C
enriched and the outer contains natural carbon (1.1 \% $^{13}$C
abundance), which are randomly distributed for both shells.}
\label{CACReview_DWCNTwirediagram}
\end{center}
\end{figure}

The inner tubes grown inside SWCNTs from peapods turned out to be a
particularly interesting system as they are remarkably defect free
which results in very long phonon life-times, i.e. very narrow
vibrational modes \cite{PfeifferPRL2003}. In addition, their smaller
diameters results in a larger energy spread, i.e. larger spectral
splitting, for diameter dependent phonon modes such as e.g. the
radial breathing mode (RBM). These two effects make the inner tubes
very suitable to study diameter dependent physics of the small
diameter tubes with precision. Here, we review how to employ the
inner tubes as $probes$ of the outer tube properties. The additional
benefit of the inner tube growth from fullerenes is that the
starting carbon source can be tailored at wish, e.g. when $^{13}$C
isotope enriched fullerenes are encapsulated inside the SWCNT host
tubes, $^{13}$C isotope enriched inner tubes are grown. In Fig.
\ref{CACReview_DWCNTwirediagram} we show the schematics of such a
DWCNT.

Here, we review the efforts to study the SWCNTs properties through
encapsulation using Raman and magnetic resonance spectroscopy. The
reviewed phenomena include the precise characterization of diameter
distribution of SWCNTs, the study of reversible hole engineering on
the SWCNTs, study of the inner tube growth mechanism with the help
of $^{13}$C isotope labeling, the study of local density of states
on the tubes using nuclear magnetic resonance (NMR) on the $^{13}$C
isotope enriched inner tubes, and the electron spin resonance (ESR)
studies of the SWCNTs using encapsulated magnetic fullerenes. This
review is organized as follows. First, we present the general
properties of DWCNTs using Raman, discuss the electronic and
vibrational properties of the inner tubes, which are the probes in
the subsequent studies. Second, we present the use of the inner
tubes to probe the host outer tube diameter distribution and to
study the opening and closing of holes on the outer tubes. Third, we
present a study on the inner tube growth mechanism using isotope
enriched carbon. Fourth, we discuss the efforts related to studying
the SWCNT properties by encapsulating magnetic fullerenes using ESR.
Fifth, we discuss the NMR results on the isotope enriched inner
tubes and in particular we present the observation of a low energy
spin-gap in the density of states of SWCNTs.

\section{Experimental methods and sample preparation}

\subsection{Sample preparation}

%\sussubsection{The starting SWCNT samples}

\begin{center}
\textbf{The starting SWCNT samples}
\end{center}

SWCNTs from different sources and prepared by different methods were
used. Commercial arc-discharge grown SWCNTs with 50 \% weight purity
(Nanocarblab, Moscow, Russia) and laser ablation prepared SWCNTs
with 10 \% weight purity (Tubes@Rice, Houston, USA) were used. The
latter material was purified through repeated steps of air oxidation
and washing in HCl. Some laser ablation prepared and purified
samples were obtained from H. Kataura. The purified samples are
usually well opened to enable fullerene encapsulation. If not,
annealing in air at 450 $^{\circ }$C for 0.5 hour makes them
sufficiently open. The HiPco samples used as reference were
purchased from CNI (Carbon Nanotechnologies Inc., Houston, USA).
Most samples were used in the form of a buckypaper, which is
prepared by filtering a suspension of SWCNTs. We found that
commercially available SWCNTs already meet a required standard in
respect of purity and quality. In addition, for the amount of
experimental work described here, reproducible samples i.e. a large
amount of SWCNTs from similar quality, were required. Commercial
samples meet this requirement, which compensates for their slightly
inferior quality compared to laboratory prepared ones.

%\subsubsection{Synthesis of peapods}

\begin{center}
\textbf{Synthesis of peapods}
\end{center}

Encapsulation of fullerenes at low temperatures inside SWCNTs
(solvent method) was performed by sonicating the fullerene and
opened SWCNT suspensions together in organic solvents following
Refs. \cite{YudasakaCPL,SimonCPL2004,Monthioux2004,BriggsJMC}. For
fullerene encapsulation at high temperatures (the vapor method), the
SWCNTs and the fullerenes were sealed under vacuum in a quartz
ampoule and annealed at 650 $^{\circ }$C for 2 hours
\cite{KatauraSM2001}. Fullerenes enter the inside of the SWCNTs at
this temperature due to their high vapor pressure that is maintained
in the sealed environment. Non-encapsulated fullerenes were removed
by dynamic vacuum annealing at the same temperature for 1 hour. High
purity fullerenes were obtained from a commercial source (Hoechst
AG, Frankfurt, Germany). The filling of SWCNTs with the fullerenes
was characterized by observing the peapod structure in
high-resolution transmission electron microscopy (HR-TEM), by x-ray
studies of the one-dimensional array of fullerenes inside the SWCNTs
and by the detection of the fullerene modes from the cages
encapsulated inside the SWCNTs using Raman spectroscopy
\cite{KatauraSM2001,PichlerPRL2001}.

%\subsubsection{Synthesis of DWCNTs}

\begin{center}
\textbf{Synthesis of DWCNTs}
\end{center}

DWCNTs were prepared by two routes: from fullerene peapods and using
chemical vapor deposition (CVD) growth technique
\cite{SimonCPL2005}. The peapods were transformed to DWCNTs by a
dynamic vacuum treatment at 1250 $^{\circ }$C for 2 hours following
Ref. \cite{BandowCPL2001}. Again, the DWCNT transformation was
followed by HR-TEM and by the observation of the DWCNT structure
factors using x-ray studies. In addition, new Raman modes emerge
after the 1250 $^{\circ}$C heat treatment particularly in a
frequency range that is clearly upshifted from the outer tube RBMs.
For the CVD DWCNT growth \cite{SimonCPL2005}, the catalyst was a
modified version of the Fe/Mo/MgO system developed by Liu \textit{et
al.} \cite{LiuCPL2004} for SWCNT synthesis.

Both kinds of DWCNTs have advantageous and disadvantageous
properties. For peapod template grown DWCNTs, the inner tube is
known to fill only up to $\sim$ 70 \% of the outer tube length
\cite{AbePRB2003}. This is the consequence of insufficient carbon in
the fullerenes: the C$_{60}$ peapods have 60 carbon atoms per 1 nm
(the lattice constant of the peapod) whereas the (9,0) inner tube
with $d=0.708$, which is representative of the most abundant 7 nm
diameter inner tube contains 36 carbon atoms per the $c_{0}=0.424$
nm lattice constant \cite{ZolyomiPRB2004}. In contrast, CVD grown
inner tubes fill up to the total length of the outer tubes, however
such samples have usually a less well defined tube diameter
distribution due to the inevitable growth of small diameter SWCNTs
and large diameter DWCNTs \cite{EndoNAT}. Peapod template grown
DWCNTs can be grown with relatively narrow diameter distribution due
to the available narrow diameter distribution of the SWCNT host
tubes. This also allows for a good control over the DWCNT diameter
as described in Ref. \cite{SimonPRB2005} and is discussed below.

%\subsubsection{Synthesis of isotope engineered DWCNTs}

\begin{center}
\textbf{Synthesis of isotope engineered DWCNTs}
\end{center}

Commercial $^{13}$C isotope enriched fullerenes (MER Corp., Tucson,
USA) were used to prepare fullerene peapods C$_{60}$,C$_{70}$@SWCNT
with enriched fullerenes. Two supplier specified grades of $^{13}$C
enriched fullerene mixtures were used: 25 and 89 \%, whose values
were slightly refined based on the Raman spectroscopy. The 25 \%
grade was
nominally C$_{60}$, and the 89 \% grade was nominally C$_{70}$ with C$_{60}$%
/C$_{70}$/higher fullerene compositions of 75:20:5 and 12:88:$<$ 1,
respectively. The above detailed standard routes were performed for
the peapod and the DWCNT productions.

\subsection{Experimental methods}

%\subsubsection{Raman spectroscopy}

\begin{center}
\textbf{Raman spectroscopy}
\end{center}

Raman spectra were measured with a Dilor xy triple spectrometer
using various lines of an Ar/Kr laser, a He/Ne laser and a tunable
Ti:sapphire and Rhodamin dye-laser in the 1.54-2.54 eV (805-488 nm)
energy range. Tunable lasers allow to record the so-called Raman map
\cite{FantiniPRL2004,TelgPRL2004} i.e. to detect the SWCNT resonance
energies through the Raman resonance enhancement \cite{KuzmanyBook},
which ultimately allows the chiral index assignment. The spectra can
be recorded in normal (NR) and high resolution (HR) mode,
respectively ($\Delta\bar{\nu}_{\text{NR}}=1.3 {\text{ cm}^{-1}}$
for blue and $\Delta\bar{\nu}_{\text{HR}}=0.4 {\text{ cm}^{-1}}$ in
the red). The samples in the form of bucky-paper are kept in dynamic
vacuum and on a copper tip attached to a cryostat, which allows
temperature variation in the 20-600 K temperature range. Raman
spectroscopy was used to characterize the diameter distribution of
the SWCNTs, to determine the peapod concentrations, and to monitor
the DWCNT transformation of the peapod samples.

%\subsection{Electron spin resonance}

\begin{center}
\textbf{Electron spin resonance}
\end{center}

The peapod and the reference SWCNT materials were mixed with the ESR
silent high purity SnO$_{2}$ in a mortar to separate the pieces of
the conducting bucky-papers. The samples were sealed under dynamic
vacuum. A typical microwave power of 10 $\mu$W and 0.01 mT magnetic
field modulation at ambient temperature were used for the
measurements in a Bruker Elexsys X-band spectrometer.

%\subsection{Nuclear magnetic resonance}

\begin{center}
\textbf{Nuclear magnetic resonance}
\end{center}

Nuclear magnetic resonance (NMR) is usually an excellent technique
for probing the electronic properties at the Fermi level of metallic
systems. The examples include conducting polymers, fullerenes, and
high temperature superconductors. However the 1.1\% natural
abundance of $^{13}$C with nuclear spin $I$=1/2 limits the
sensitivity of such experiments. As a result, meaningful NMR
experiments have to be performed on $^{13}$C isotope enriched
samples. NMR data were taken with the samples sealed in quartz tubes
filled with a low pressure of high purity Helium gas
\cite{SimonPRL2005}. We probed the low frequency spin dynamics (or
low energy spin excitations, equivalently) of the inner-tubes using
the spin lattice relaxation time, $T_{1}$, defined as the
characteristic time it takes the $^{13}$C nuclear magnetization to
recover after saturation. The signal intensity after saturation,
$M(t)$, was deduced by integrating the fast Fourier transform of
half the spin-echo for different delay times, $t$.

\section{Results and discussion}

\subsection{Inner tubes in DWCNTs as local probes}

\subsubsection{Electronic and vibrational properties of DWCNTs}

\begin{figure}
\begin{center}
\includegraphics[angle=270,width=1\linewidth]{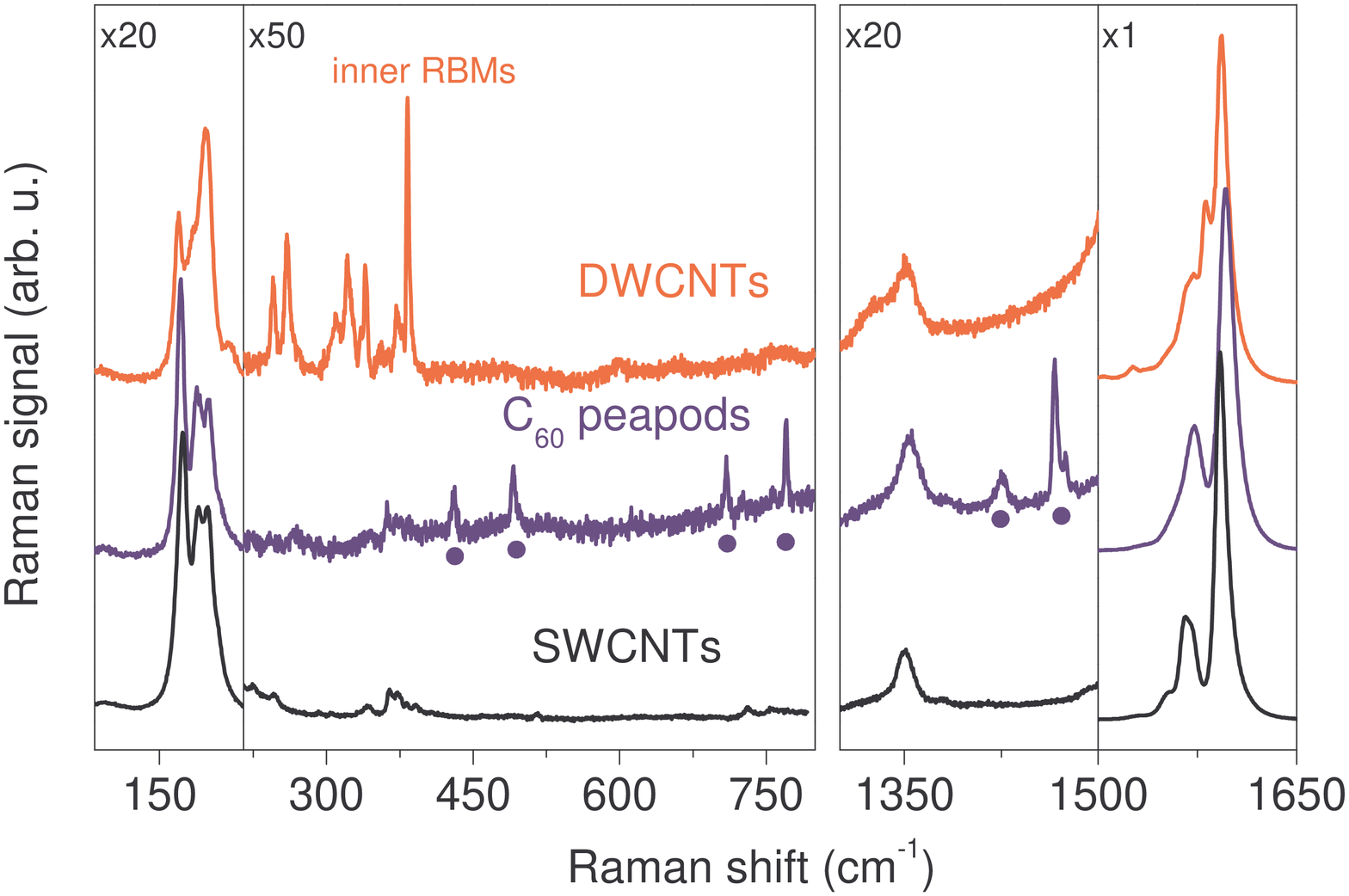}
\caption{Transformation of fullerene peapods to DWCNTs as followed
with Raman spectroscopy at 496.5 nm laser excitation and 90 K. The
SWCNT Raman spectra (lower curve) is shown as reference. The
fullerene related peapod modes (dots) in the middle curve disappear
upon the heat treatment. Note the sharp RBMs appearing in the
250-450 cm$^{-1}$ for the DWCNT sample.}
\label{CACreview_DWCNTtransformation}
\end{center}
\end{figure}

Encapsulating fullerenes and transforming them into inner tubes by
the high temperature annealing process \cite{BandowCPL2001} provides
a unique opportunity to study the properties of the host outer
tubes. In Fig. \ref{CACreview_DWCNTtransformation} we show the
evolution of the SWCNT Raman spectrum upon C$_{60}$ fullerene
encapsulation and the DWCNT transformation after Ref.
\cite{PfeifferPRL2003}. The series of sharp modes in the peapod
spectrum, which are related to the encapsulated fullerenes
\cite{PichlerPRL2001}, disappear upon the heat treatment and a
series of sharp modes appear in the 250-450 cm$^{-1}$ spectral
range. The presence of inner tubes after this protocol have been
independently confirmed by HR-TEM \cite{LuzziCPL2000}. The small
diameter tubes with $d \sim$ 0.7 nm would have RBM modes in the
$\sim$ 250-450 cm$^{-1}$ spectral range, which clarifies the
identification of these modes. The identification of the inner tube
RBMs is possible due to the strong $d$ dependence of this Raman mode
\cite{Kuerti:PhysRevB58:R8869:(1998)}. Assignment of less diameter
dependent modes such as the G mode \cite{DresselhausTubes} to inner
and outer tubes are more difficult although a number of small
intensity new modes are observed for the DWCNT sample in Fig.
\ref{CACreview_DWCNTtransformation}. It is shown in Section
\ref{isotope_labeled} that unambiguous assignment can be given with
the help of selective isotope enrichment of the inner walls.

\begin{figure}
\begin{center}
\includegraphics[width=0.9\linewidth]{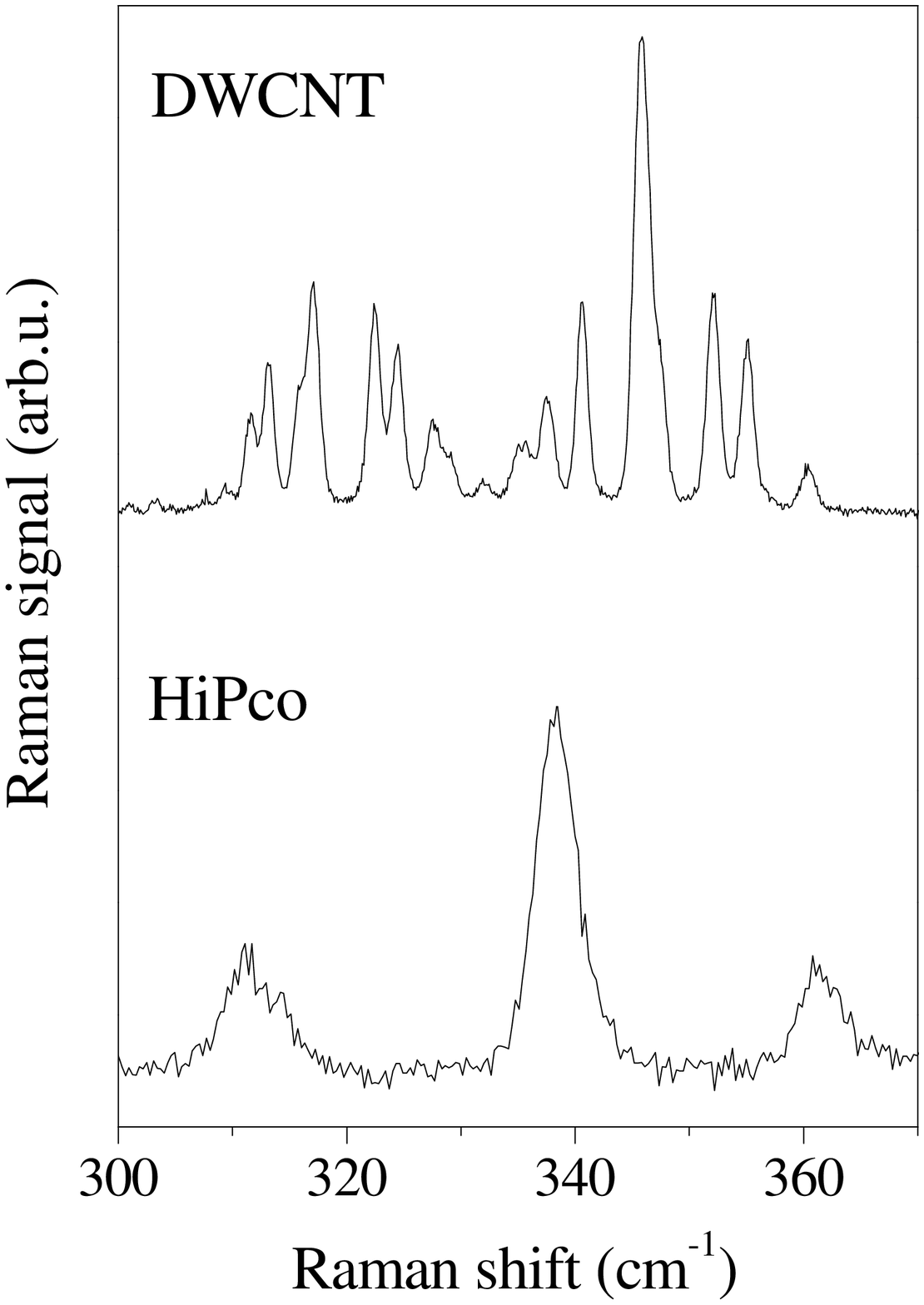}
\caption{Raman spectra of the RBMs in DWCNT and HiPco (SWCNT)
samples at 594 nm laser excitation and 90 K in the high resolution
spectrometer mode.} \label{CACReview_DWCNTHR}
\end{center}
\end{figure}

A variety of additional information can be gained about the inner
tube properties when their RBMs are studied using the additive mode,
i.e. high-resolution of the Raman spectrometer. In Fig.
\ref{CACReview_DWCNTHR}, we show the inner tube RBMs at 90 K with
high-resolution in comparison with an SWCNT sample with similar tube
diameter prepared by the HiPco process. Three striking observations
are apparent in the comparison of the two spectra: i) there are a
larger number of inner tube RBMs than geometrically allowed and they
appear to cluster around the corresponding modes in the SWCNT
sample, ii) the inner tube RBMs are on average an order of magnitude
narrower than the SWCNT RBMs in the HiPco sample
\cite{PfeifferPRL2003} and iii) the Raman intensity of the inner
tubes is large in view of the $\sim$ 3 times less number of carbon
atoms on them \cite{PfeifferPRB2004}. Points ii) and iii) are
explained by the long phonon and quasi-particle life-times of inner
tubes which are discussed further below.

\begin{figure}
\begin{center}
\includegraphics[width=\linewidth]{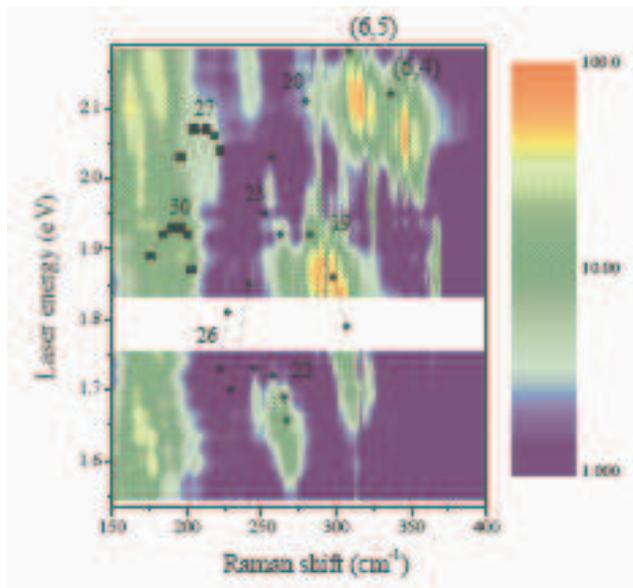}
\caption{Raman map of DWCNTs. Circles and squares are the
$E_{22}^{\text{s}}$ and $E_{11}^{\text{m}}$ peaks as measured in a
HiPco sample \cite{FantiniPRL2004}, respectively. The family numbers
and the chiral indexes for the (6,5) and (6,4) tubes are indicated.
Dashed lines join chiralities in the same family. Laser excitation
was not available in the missing area. Reprinted figure with
permission from Ref. \cite{PfeifferPRB2005b}, R. Pfeiffer \textit{et
al.} Phys. Rev. B \textbf{72}, 161404 (2005). Copyright (2005) by
the American Physical Society.} \label{CACReview_DWCNTRamanMapFull}
\end{center}
\end{figure}

Observation i), i.e. the clustering behavior of the observed inner
tube RBMs around SWCNT RBMs, is further evidenced in energy
dispersive Raman measurements. In Fig.
\ref{CACReview_DWCNTRamanMapFull}, we show the Raman map for the
DWCNTs from Ref. \cite{PfeifferPRB2005b}. The advantage of studying
Raman maps is that the optical transition energies are also
contained in addition to the Raman shifts. These two quantities
uniquely identify the chirality of a nanotube
\cite{DresselhausTubes,DresselhausTubesNew,KatauraSM1999}. The
analogous Raman map for HiPco SWCNTs were measured by Fantini
\textit{et al.} \cite{FantiniPRL2004} and Telg \textit{et al.}
\cite{TelgPRL2004}. Their results are also shown in Fig.
\ref{CACReview_DWCNTRamanMapFull}) with squares and circles for
metallic and semiconducting tubes, respectively. It turns out that
family patterns with $2n+m=$ const can be identified for which the
tube resonance energies and Raman shifts are closely grouped
together \cite{Bachilo:Science298:2361:(2002)}. The comparison of
the HiPco results and the DWCNT Raman map confirms the above
statement, i.e. that a number of inner tube modes are observed for
the DWCNT where only a few (or one) SWCNT chirality is present. This
is best seen for the (6,5) and (6,4) chiralities which are well
resolved from other modes.

\begin{figure}
\begin{center}
\includegraphics[width=\linewidth]{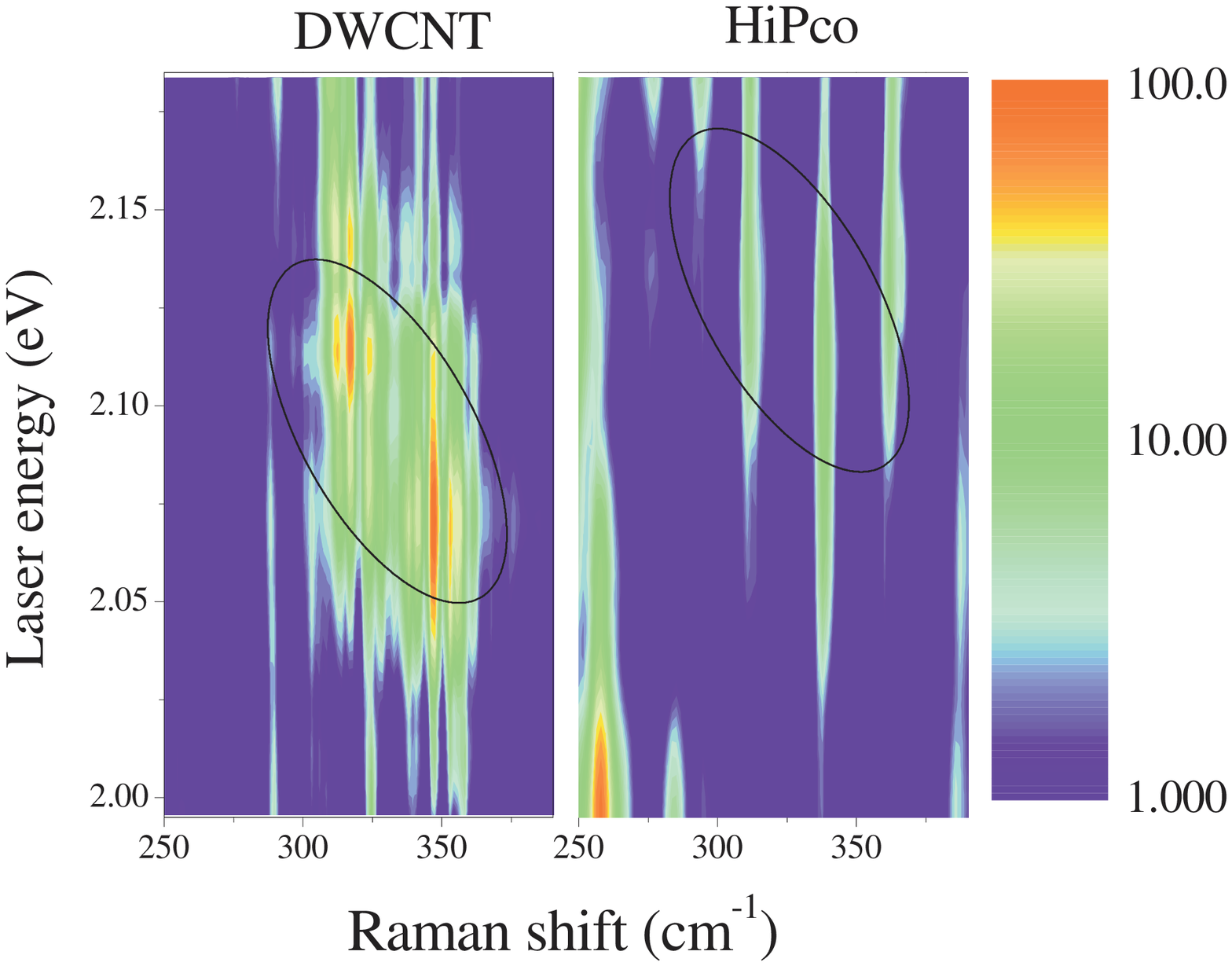}
\caption{Raman map comparison of the (6,5) and (6,4) tube RBM ranges
for DWCNT and SWCNT (HiPco) samples. Ellipsoids indicate the
corresponding tube modes. Note the progressive transition energy
downshift for the split components of the inner tubes and the 30 meV
transition energy difference between the two kinds of samples, which
are discussed in the text. } \label{CACReview_DWCNTHIPCORamanMaps}
\end{center}
\end{figure}

In Fig. \ref{CACReview_DWCNTHIPCORamanMaps}, we show the Raman maps
for the two samples near the energy and Raman shift regions for the
(6,5) and (6,4) tube modes
\cite{Bachilo:Science298:2361:(2002),FantiniPRL2004,TelgPRL2004}.
The comparison of the Raman maps of the two kinds of samples shows
that the corresponding tube modes are split into up to 15 components
for the inner tube RBMs. This is explained by the inner-outer tube
interaction in the DWCNT samples: an inner tube with a particular
chirality can be grown in outer tubes with different diameters
(chiralities). The varying inner-outer tube spacing can give rise to
a different Raman shift for the split components. The large number
of split components is a surprising result as it is expected that an
inner tube with a given diameter is grown in maximum 1-2 outer tubes
where its growth is energetically preferred.

To further prove the origin of the splitting and to quantify this
effect, model calculations on the inner-outer tube interactions were
performed \cite{PfeifferPRB2005b,PfeifferEPJB2004} following the
continuum model of Popov and Henrard
\cite{Popov:PhysRevB65:235415:(2002)}. These calculations showed
that the interaction of inner and outer tubes can gives rise to a
shift in the inner tube RBM frequency up to 30 cm$^{-1}$.

\subsubsection{Phonon and quasi-particle life-times in DWCNTs}

\begin{figure}
\begin{center}
\includegraphics[width=0.9\linewidth]{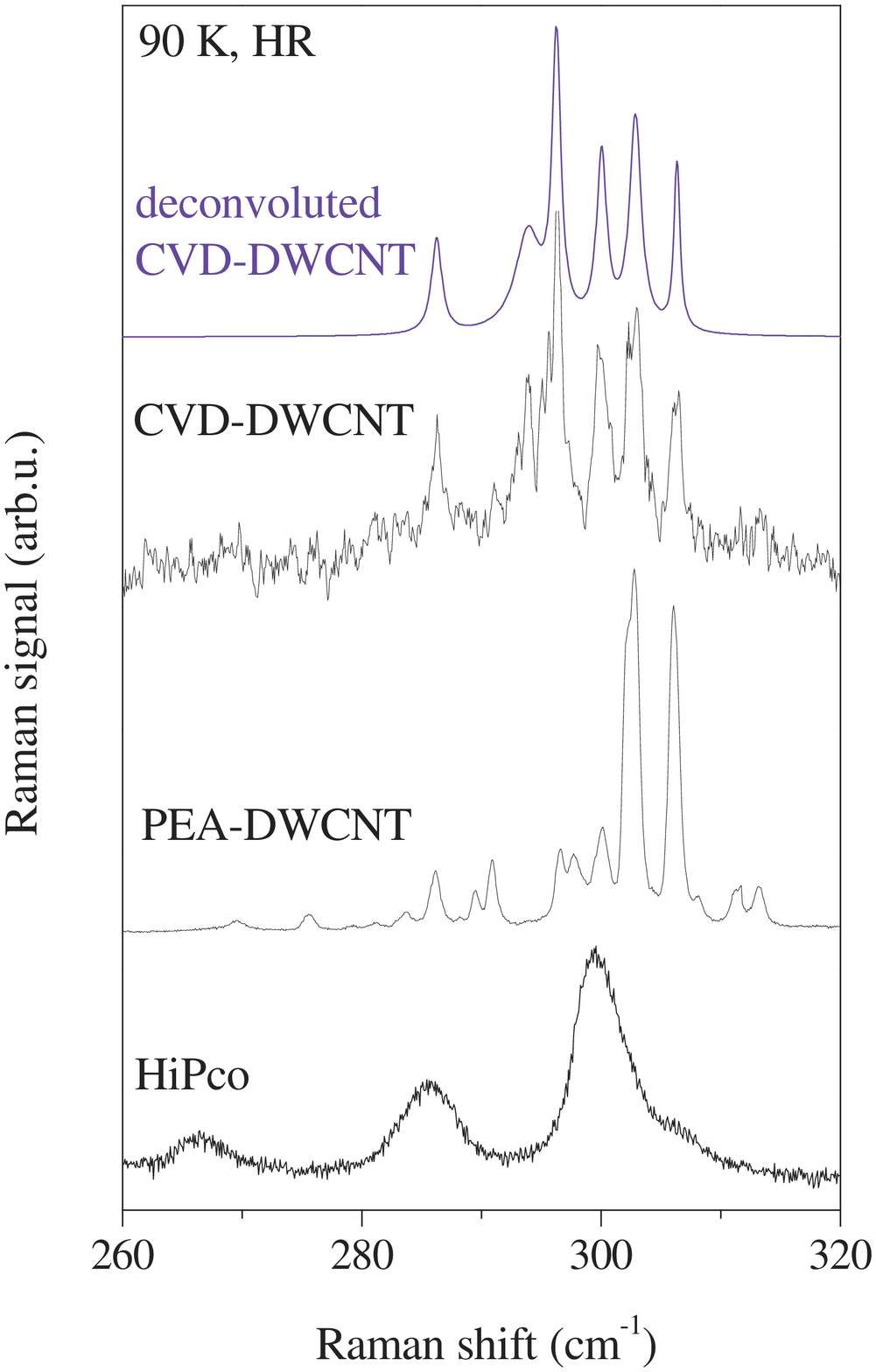}
\caption{High resolution Raman spectra taken at 676 nm laser
excitation and 90 K on the CVD- and PEA-DWCNT and an SWCNT reference
(HiPco) sample. The deconvoluted spectrum is also shown for the
CVD-DWCNT sample. The narrow line-widths indicate the long RBM
phonon life-times of the inner tubes in both DWCNT materials.
Reprinted figure with permission from Ref. \cite{SimonCPL2005}, F.
Simon \textit{et al.} Chem. Phys. Lett. \textbf{413}, 506 (2005).
Copyright (2005) by Elsevier.} \label{CACReview_CVDDWCNT}
\end{center}
\end{figure}

Now, we turn to discussion of the observed very narrow line-widths
of the RBMs. This is the most important property of the inner tube
RBMs, which will be exploited throughout in this work. Intrinsic
line-widths can be determined by deconvoluting the experimental
spectra with a Voigtian fit, whose Gaussian component describes the
spectrometer resolution and the Lorentzian gives the intrinsic
line-width. The Lorentzian component for some inner tube RBMs is as
small as 0.4 cm$^{-1}$ \cite{PfeifferPRL2003}, which is an order or
magnitude smaller than the values obtained for isolated individual
tubes in a normal SWCNT sample
\cite{Jorio:PhysRevLett86:1118:(2001)}. The narrow line-widths, i.e.
long phonon life-times of the inner tube RBMs was originally
associated to the perfectness of the inner tubes grown from the
peapod templates \cite{PfeifferPRL2003}. It was found, however, that
inner tubes in chemical vapor deposition (CVD) grown DWCNTs have
similarly small line-widths \cite{SimonCPL2005}. In Fig.
\ref{CACReview_CVDDWCNT}, the high resolution spectra for the inner
tube RBMs in CVD and peapod template grown DWCNTs is shown. This
suggests, that the tube environment plays an important role in the
magnitude of the observable RBM line-width.

The tube-tube interactions have been shown to give rise to up to
$\approx$ 30 cm$^{-1}$ extra shift to the RBMs
\cite{PfeifferPRB2005b}. The principal difference between SWCNTs and
inner tubes in DWCNTs (irrespective whether these are CVD or peapod
template grown) is the different surrounding of a small diameter
SWCNT with a given chirality: for the SWCNT sample, each tube is
surrounded by the ensemble of other SWCNTs. For a close packed
hexagonal bundle structure \cite{Thess:Science273:483:(1996)}, this
involves 6 nearest neighbors with random chiralities. This causes an
inhomogeneous broadening of the RBMs. However, the nearest-neighbor
of an inner tube with a given chirality is an outer tube also with a
well defined chirality. A given inner tube can be grown in several
outer tubes with different diameters, however the chiralities of an
inner-outer tube pair is always well defined, therefore the nearest
neighbor interaction acting on an inner tube is also well-defined.

\begin{figure}[tbp]
\begin{center}
\includegraphics[width=1.0\hsize]{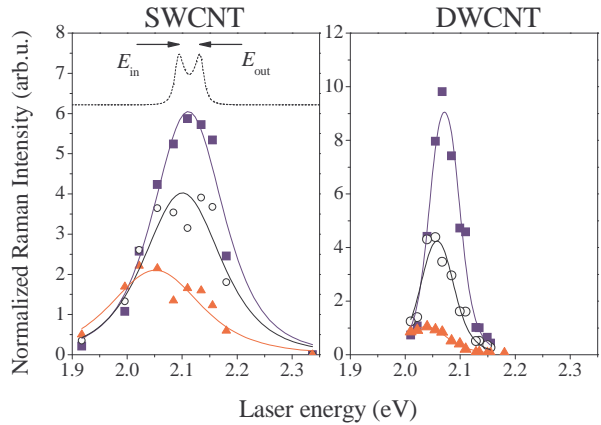}
\caption{Raman resonance profile for the (6,4) tubes in the SWCNT
(CoMoCat) and DWCNT samples, $\blacksquare$: 80 K, $\bigcirc$: 300
K, $\blacktriangle$: 600 K. Solid curves show fits with the RRS
theory. Dashed curve is a simulation for the 80 K SWCNT data with
$\Gamma=10$ meV. Arrows indicate the incoming and outgoing resonance
energies. Note the much narrower widths for the DWCNT sample.
Reprinted figure with permission from Ref. \cite{SimonPRB2006}, F.
Simon \textit{et al.} Phys. Rev. B \textbf{74}, 121411(R) (2006).
Copyright (2006) by the American Physical Society. }
\label{CACReview_DWCNTCOMOCAT_ERG_PROFILE}
\end{center}
\end{figure}

In addition to the long phonon life-times of inner tubes, the
life-time of optical excitations, i.e. the life-time of the
quasi-particle associated with the Raman scattering is unexpectedly
long. To demonstrate this, we compare the resonant Raman scattering
data for an inner tube and a SWCNT with the same chirality following
Ref. \cite{SimonPRB2006}. In Fig.
\ref{CACReview_DWCNTCOMOCAT_ERG_PROFILE} we show the energy profile
of the resonant Raman scattering at some selected temperatures for
two 6,4 tube modes: one is an inner tube in a DWCNT sample, the
other is a SWCNT in a CoMoCat sample. Such energy profiles are
obtained by taking an energy (vertical) cross section of a Raman map
such as shown in Fig. \ref{CACReview_DWCNTHIPCORamanMaps}. The Raman
intensities for a given excitation energy were obtained by fitting
the spectra with Voigtian curves for the tube modes, whose Gaussian
component accounts for the spectrometer resolution and whose
Lorentzian for the intrinsic line-width. For the DWCNT sample, the
strongest (6,4) inner tube component at 347 cm$^{-1}$ and for the
SWCNT CoMoCat sample the (6,4) tube mode at 337 cm$^{-1}$ is shown.
The temperature dependent resonant Raman data can be fitted with the
conventional resonance Raman theory for Stokes Raman modes
\cite{KuzmanyBook,KuzmanyEPJB}:

\begin{eqnarray}
%\phantom{mmmmmmmmmmmmmmmmmmmm} \nonumber \\
I(E_{\text{l}})  =M_{\text{eff}}^{4}\left\vert
\frac{\left(E_{\text{l}}-E_{\text{ph}}\right)^4\left(n_\text{BE}(E_{\text{ph}})+1\right)}{\left(
E_{\text{l}}-E_{\text{ii}}-i\Gamma \right) \left(
E_{\text{l}}-E_{\text{ph}}-E_{\text{ii}}-i\Gamma \right)
}\right\vert ^{2} \label{CACReview_Resonance_Raman}
\end{eqnarray}

\noindent Here, the electronic density of states of SWCNTs is
assumed to be a Dirac function and the effective matrix element,
$M_{\text{eff}}$, describing the electron-phonon interactions is
taken to be independent of temperature and energy. $E_{\text{l}}$,
$E_{22}$ and $E_{\text{ph}}$ are the exciting laser, the optical
transition and the phonon energies, respectively.
$n_\text{BE}(E_{\text{ph}})=(\exp(
E_{\text{ph}}/\text{k}_{\text{B}}T)-1)^{-1}$ is the Bose-Einstein
function and accounts for the thermal population of the vibrational
state \cite{KuzmanyBook} and $n_\text{BE}(E_{\text{ph}})+1$ changes
a factor $\sim$ 2 between 80 and 600 K. The temperature dependence
of $E_{\text{ph}}$ is $\sim$ 1 \% for the studied temperature range
\cite{AjayanPRB2002} thus it can be neglected. The first and second
terms in the denominator of Eq. \ref{CACReview_Resonance_Raman}
describe the incoming and outgoing resonances, respectively and are
indicated on a simulated curve by arrows in Fig.
\ref{CACReview_DWCNTCOMOCAT_ERG_PROFILE}. These are separated by
$E_{\text{ph}}$. This means the apparent width of the resonance
Raman data does not represent $\Gamma$.

Clearly, the resonance width is always smaller for the DWCNT than
for the SWCNT sample. In other words, the life-time of the optically
excited quasi-particle is longer lived for the DWCNT. The
quasi-particle life-time is an important parameter for the
application of carbon nanotubes in optoelectronic devices
\cite{AvourisPRL2004,AvourisPRL2005}. As a result, DWCNTs appear to
be superior in this respect than their one-walled counterparts.

\subsubsection{Probing the SWCNT diameter distribution through inner tube growth}

As discussed above, the Raman spectra of inner tubes have several
advantages compared to that of the outer tubes: i) their RBMs have
about a factor 2 times larger splitting due to the smaller
diameters, ii) the line-widths are about 10 times narrower. The
larger spectral splitting and narrower line-widths of the inner tube
RBMs enable to characterize the inner tube diameter distribution
with a spectral resolution that is about $20$\ times larger as
compared to the analysis on the outer tubes. To prove that studying
the inner tubes can be exploited for the study of outer ones, here
we show that there is a one-to-one correspondence between the inner
and outer tube diameter distributions following Ref.
\cite{SimonPRB2005}.

\begin{figure}[tbp]
\begin{center}
\includegraphics[width=0.85\hsize]{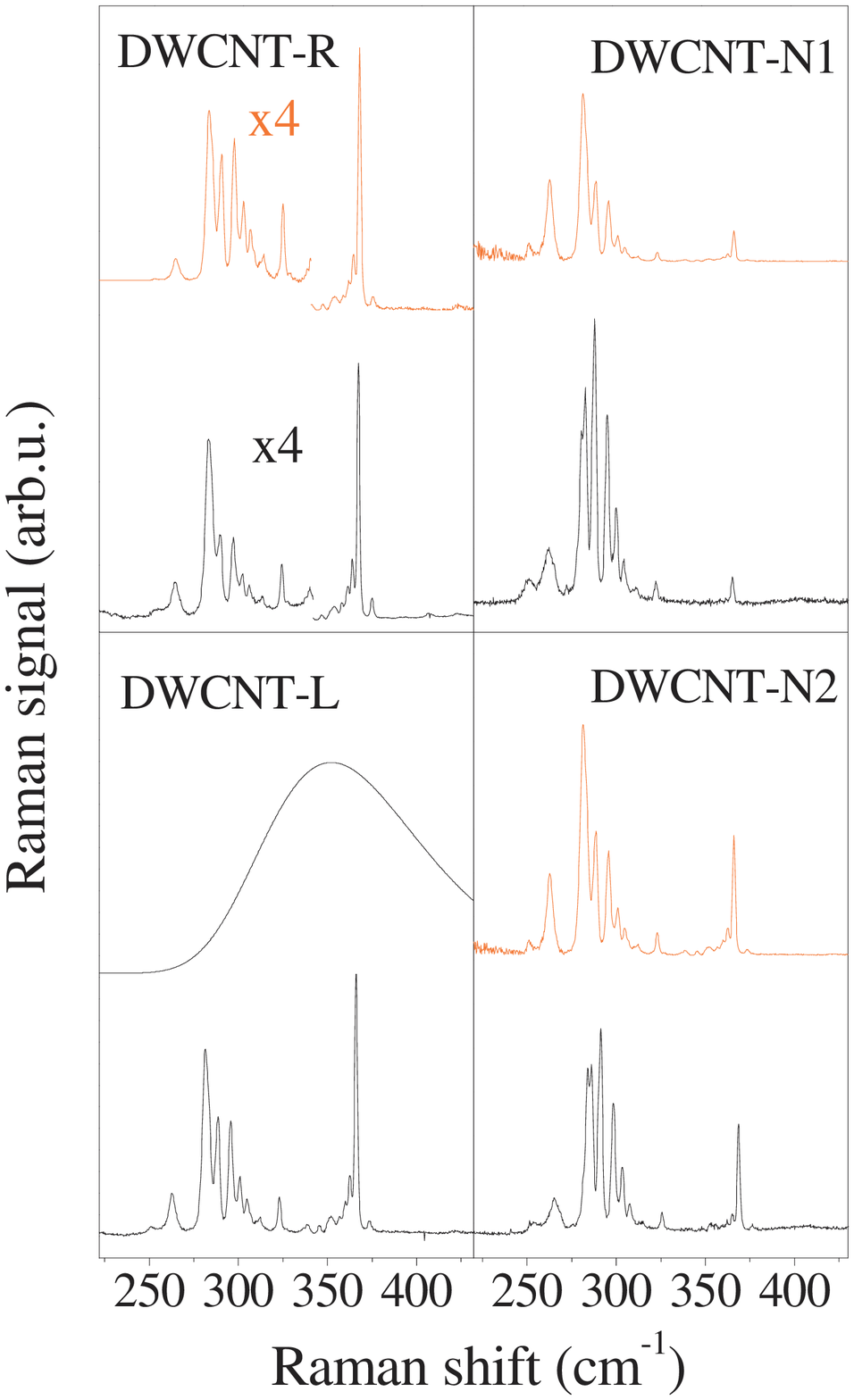}
\caption{As measured Raman spectra of the inner nanotube RBMs for
four DWCNT samples (lower curves in each quarter) at 647 nm laser
excitation. The upper spectra (shown in red) are "smart-scaled" from
the lower left spectrum. The Gaussian diameter distribution is shown
for the DWCNT-L sample. Reprinted figure with permission from Ref.
\cite{SimonPRB2005}, F. Simon \textit{et al.} Phys. Rev. B
\textbf{71}, 165439 (2005). Copyright (2005) by the American
Physical Society.} \label{CACReview_FourSamples}
\end{center}
\end{figure}

In Fig. \ref{CACReview_FourSamples}, we compare the inner tube RBM
Raman spectra for four different DWCNT materials based on SWCNTs
with different diameters and produced with different methods. The
SWCNTs were two arc-discharge grown SWCNTs (SWCNT-N1 and N2) and two
laser ablation grown tubes (SWCNT-R and SWCNT-L). The diameter
distributions of the SWCNT materials were determined from Raman
spectroscopy \cite{KuzmanyEPJB} giving $d_{\text{N}1}=$ 1.50 nm,
$\sigma _{\text{N1}}$ = 0.10 nm, $d_{\text{N}2}=$ 1.45 nm, $\sigma
_{\text{N}1}$ = 0.10 nm, $d_{\text{R}}=$ 1.35 nm, $\sigma
_{\text{R}}$ = 0.09 nm, and $d_{\text{L}}=$ 1.39 nm, $\sigma
_{\text{L}}$ = 0.09 nm for the mean diameters and the variances of
the distributions, respectively.

The spectra shown are excited with a 647 nm laser that is
representative for excitations with other laser energies. The RBMs
of all the observable inner tubes, including the split components
\cite{PfeifferPRL2003}, can be found at the same position in all
DWCNT samples within the $\pm $0.5 cm$^{-1}$ experimental precision
of the measurement for the whole laser energy range studied. This
proves that vibrational modes of DWCNT samples are robust against
the starting material.

As the four samples have different diameter distributions, the
overall Raman patterns look different. However, scaling the patterns
with the ratio of the distribution functions ("smart-scaling")
allows to generate the overall pattern for all systems, starting
from e.g. DWCNT-L in the bottom-left corner of Fig.
\ref{CACReview_FourSamples}. It was assumed that the inner tube
diameter distributions follow a Gaussian function with a mean
diameter 0.72 nm smaller than those of the outer tubes following
Ref. \cite{AbePRB2003} and with the same variance as the outer
tubes. The empirical constants from Ref. \cite{KrambergerPRB2003}
were used for the RBM mode Raman shift versus inner tube diameter
expression. The corresponding Gaussian diameter distribution of
inner tubes is shown for the DWCNT-L sample in Fig.
\ref{CACReview_FourSamples}. A good agreement between the
experimental and simulated patterns for the DWCNT-R sample is
observed. A somewhat less accurate agreement is observed for the
DWCNT-N1, N2 samples, which may be related to the different growth
method: arc discharge for the latter, as compared to laser ablation
for the R and L samples. The observed agreement has important
consequences for the understanding of the inner tube properties. As
a result of the photoselectivity of the Raman experiment, it proves
that the electronic structure of the inner tubes is identical in the
different starting SWCNT materials.

The scaling of the inner tube Raman spectra with the outer tube
distribution shows that the inner tube abundance follows that of the
outer ones. This agrees with the findings of x-ray diffractomery on
DWCNTs \cite{AbePRB2003} and is natural consequence of the growth of
inner tubes inside the outer tube hosts.

\subsubsection{Studying the reversible hole engineering using DWCNTs}

Soon after the discovery of the peapods \cite{SmithNAT}, it was
recognized \cite{KatauraSM2001,SmithCPL2000} that opening the SWCNTs
by oxidation in air or by treating in acids is a prerequisite for
good filling. Good filling means a macroscopic filling where the
peapods are observable not only by local microscopic means such as
HR-TEM but also by spectroscopy such as Raman scattering. On the
other hand, a heat treatment around 1000 $^{\circ }$C was known to
close the openings which results in a low or no fullerene filling.
It was also shown that the geometrically possible maximum of filling
can be achieved when purified SWCNTs were subject to a 450 $^{\circ
}$C heat treatment in flowing oxygen \cite{LiuPRB2002}. However,
these studies have concerned the overall fullerene filling, with no
knowledge on the precise dependence on the thermal treatment or tube
diameter specificity.

The high diameter and chirality sensitivity of Raman spectroscopy
for the inner tubes allows to study the behavior of tube openings
when subject to different treatments. More precisely, openings which
allow fullerenes to enter the tubes can be studied. This is achieved
by studying the resulting inner tube RBM pattern when the outer tube
host was subject to some closing or opening treatments prior to the
fullerene encapsulation \cite{HasiJNN}. Annealing of as purchased or
opened tubes was performed at various temperatures between 800
$^{\circ }$C and 1200 $^{\circ }$C in a sealed and evacuated quartz
tube at a rest gas pressure of 10$^{-6}$ mbar. Opening of the tubes
was performed by exposure to air at various temperatures between 350
$^{\circ }$C and 500 $^{\circ }$C.

\begin{figure}[tbp]
\begin{center}
\includegraphics[width=0.8\hsize]{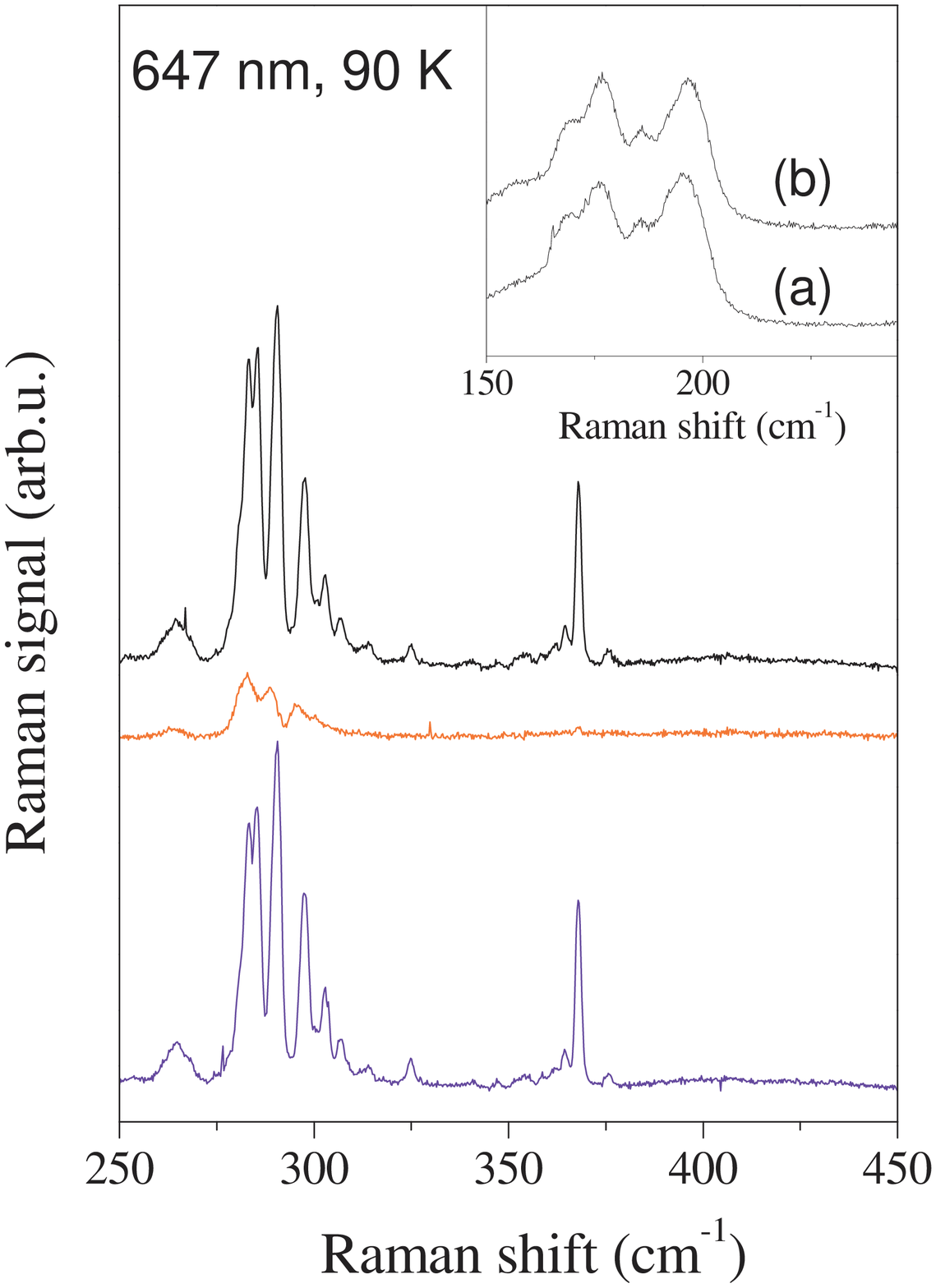}
\caption{Raman spectra in the spectral range of the inner shell tube
RBM for nanotubes after special pre-treatment. Bottom: after filling
the tubes with C$_{60}$ and standard transformation; Center: after
annealing the tubes at 1000 $^{\circ }$C, filling with C$_{60}$ and
standard transformation; Top: after re-opening the annealed samples,
filling with C$_{60}$, and standard transformation. All spectra
recorded at 90 K and $\lambda=647$ nm. Insert: the RBM of the outer
tubes before (a) and after (b) annealing at 1000 $^{\circ }$C.
Reprinted figure with permission from Ref. \cite{HasiJNN}, F. Hasi
\textit{et al.} J. Nanosci. Nanotechn. \textbf{5}, 1785 (2005).
Copyright (2005) by the American Scientific Publishers.}
\label{CACReview_CloseOpen}
\end{center}
\end{figure}

Figure \ref{CACReview_CloseOpen} shows the Raman response of tubes
after the standardized DWCNT transformation conditions but different
pre-treatment. Only the spectral range of the inner tube is shown in
the main part of the figure. The spectrum at the center was recorded
under identical conditions but the SWCNT was pre-annealed before the
standardized filling and standardized transformation. Almost no
response from inner shell tubes is observed for this material, which
means no fullerenes had entered the tubes: the tubes were very
efficiently closed by the annealing process. The insert in Fig.
\ref{CACReview_CloseOpen} depicts the RBM response from the outer
tubes before and after annealing. The two spectra are almost
identical, which proves that no outer tube coalescence had occurred
at the temperature applied. The spectrum at the top in Fig.
\ref{CACReview_CloseOpen} was recorded after reopening the annealed
tubes at 500 $^{\circ }$C on air and standard filling and
transformation. The spectra derived from the pristine and from the
reopened tubes are identical in all details. This means no dramatic
damages by cutting a large number of holes into the sidewalls have
happened. Consequently, the sidewalls of the tubes remain highly
untouched by the opening process. Thus, it is suggested that
fullerenes enter the tubes through holes at the tube ends.

\subsection{Growth mechanism of inner tubes studied by isotope labeling}
\label{isotope_labeled}

The growth of inner tubes from fullerenes raises the question,
whether the fullerene geometry plays an important role in the inner
tube growth or it acts as a carbon source only. Theoretical results
suggest the earlier possibility \cite{SmalleyPRL2002,TomanekPRB}. In
addition, it needs clarification whether carbon exchange occurs
between the two tube walls. Here, we review $^{13}$C isotope labeled
studies aimed at answering these two open questions. $^{13}$C is a
naturally occurring isotope of carbon with 1.1 \% abundance. In
general, isotope substitution provides an important degree of
freedom to study the effect of change in phonon energies while
leaving the electronic properties unaffected. This has helped to
unravel phenomena such as e.g. the phonon-mediated superconductivity
\cite{BCS}.

First, we discuss the inner tube growth from isotope labeled
fullerenes \cite{SimonPRL2005}, and second we present the growth of
inner tubes from isotope labeled organic solvents
\cite{SimonCPL2006}.

Commercial $^{13}$C isotope enriched fullerenes with two different
enrichment grades were used to grow isotope enriched inner tubes.
Fullerene encapsulation \cite{KatauraSM2001} and inner tube growth
was performed with the conventional methods
 \cite{BandowCPL2001}. This results in a compelling isotope engineered system:
double-wall carbon nanotubes with $^{13}$C isotope enriched inner
walls and outer walls containing natural carbon \cite{SimonPRL2005}.

\begin{figure}[tbp]
\begin{center}
\includegraphics[width=0.9\hsize]{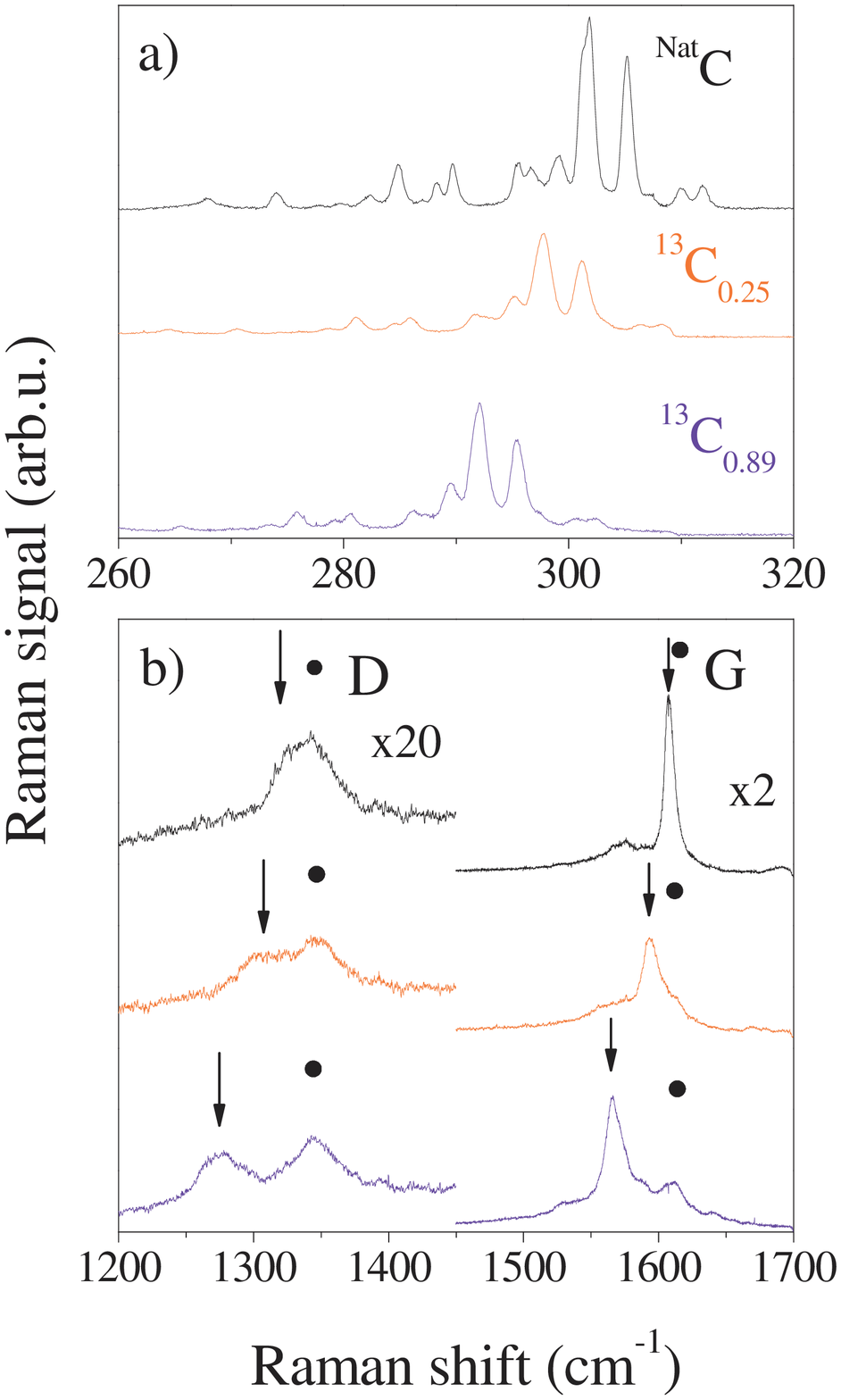}
\caption{Raman spectra of DWCNTs with natural carbon and $^{13}$C
enriched inner tubes at 676 nm laser excitation and 90 K. The inner
tube RBM (a) and D and G mode spectral ranges (b) are shown. Arrows
and filled circles indicate the D (left) and G (right) modes
corresponding to the inner and outer tubes, respectively. Reprinted
figure with permission from Ref. \cite{SimonPRL2005}, F. Simon
\textit{et al.} Phys. Rev. Lett. \textbf{95}, 017401 (2005).
Copyright (2005) by the American Physical Society.}
\label{CACReview_13DWCNT_RBM}
\end{center}
\end{figure}

In Fig. \ref{CACReview_13DWCNT_RBM}a, we show the inner tube RBM
range Raman spectra for a natural DWCNT and two DWCNTs with
differently enriched inner walls, 25 \% and 89 \%. These two latter
samples are denoted as $^{13}$C$_{25}$- and $^{13}$C$_{89}$-DWCNT,
respectively. The inner wall enrichment is taken from the nominal
enrichment of the fullerenes used for the peapod production, whose
value is slightly refined based on the Raman data. An overall
downshift of the inner tube RBMs is observed for the $^{13}$C
enriched materials accompanied by a broadening of the lines. The
downshift is clear evidence for the effective $^{13}$C enrichment of
inner tubes. The magnitude of the enrichment and the origin of the
broadening are discussed below.

The RBM lines are well separated for inner and outer tubes due to the $\nu _{%
\text{RBM}}\propto 1/d$ relation and a mean inner tube diameter of $d \sim $%
\ 0.7 nm \cite{AbePRB2003,SimonPRB2005}. However, other vibrational
modes such as the defect induced D and the tangential G modes
strongly overlap for inner and outer tubes. Arrows in Fig.
\ref{CACReview_13DWCNT_RBM}b indicate a gradually downshifting
component of the observed D and G modes. These components are
assigned to the D and G modes of the inner tubes. The sharper
appearance of the inner tube G mode, as compared to the response
from the outer tubes, is related to the excitation of semiconducting
inner tubes and metallic outer tubes
\cite{PfeifferPRL2003,SimonPRB2005}.

The shifts for the RBM, D and G modes can be analyzed for the two
grades of enrichment. The average value of the relative shift for
these modes was found to be $\left( \nu _{0}-\nu \right) /\nu
_{0}=0.0109(3)$ and $0.0322(3)$ for the $^{\text{13}}$C$_{0.25}$-
and $^{\text{13}}$C$_{0.89}$-DWCNT samples, respectively. Here, $\nu
_{0}$ and $\nu $ are the Raman shifts of the same inner tube mode in
the natural carbon and enriched materials, respectively. In the
simplest continuum model, the shift originates from the increased
mass of the inner tube walls. This gives $\left( \nu _{0}-\nu
\right) /\nu _{0}=1-\sqrt{\frac{12+c_{0}}{12+c}}$, where $c$ is the
concentration of the $^{13}$C enrichment on the inner tube, and
$c_{0}=0.011$ is the natural abundance of $^{13}$C in carbon. The
resulting values of $c$ are $0.277(7)$ and $0.824(8)$ for the 25 and
89 \% samples, respectively.

\begin{figure}[tbp]
\begin{center}
\includegraphics[width=0.8\hsize]{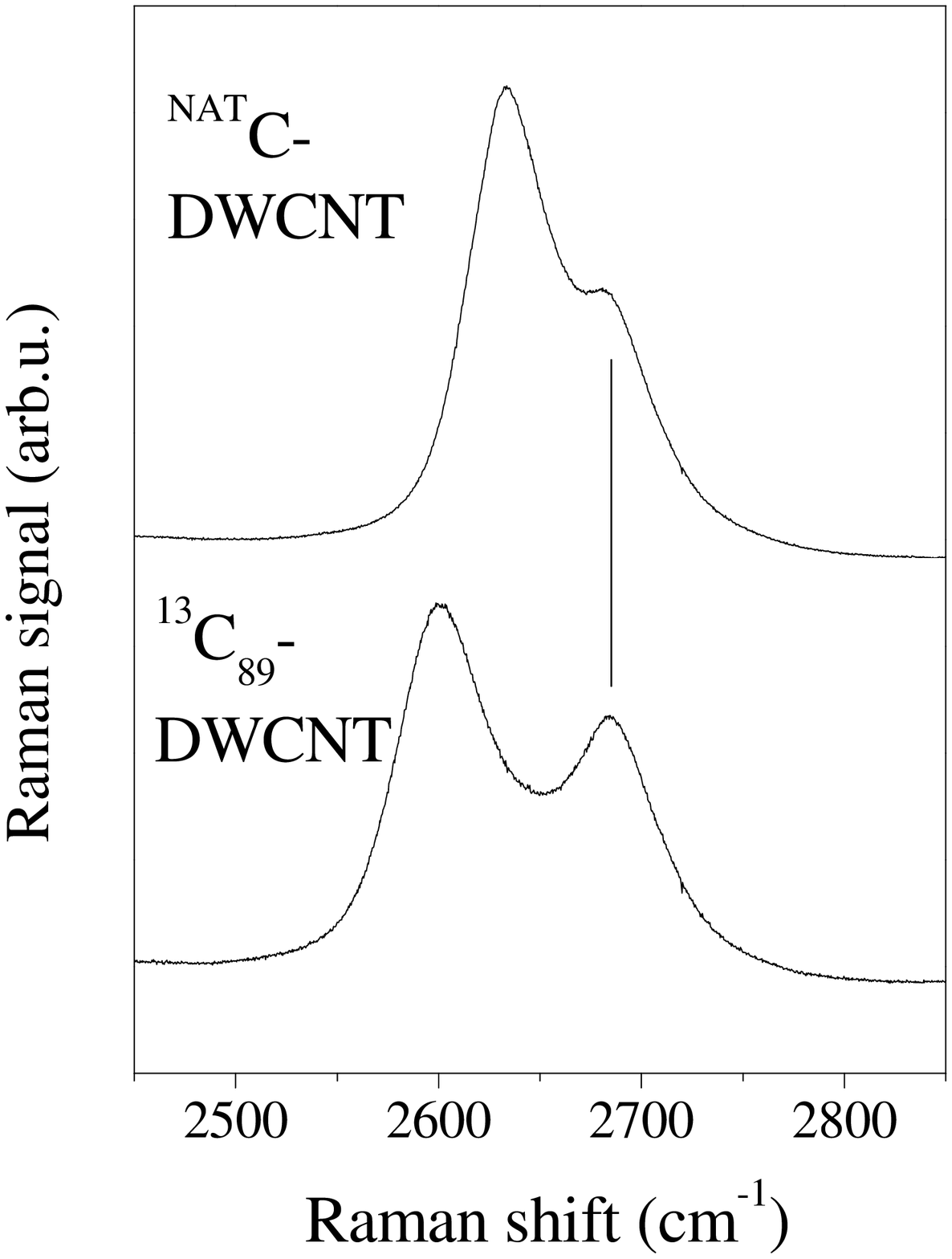}
\caption{G' spectral range of DWCNTs with natural carbon and
$^{13}$C enriched inner walls with 515 nm laser excitation. Note the
unchanged position of the outer tube G' mode indicated by a vertical
line.} \label{CACReview_13DWCNT_Dst}
\end{center}
\end{figure}

The growth of isotope labeled inner tubes allows to address whether
carbon exchange between the two walls occurs during the inner tube
growth. In Fig. \ref{CACReview_13DWCNT_Dst}, we show the G' spectral
range for DWCNTs with natural carbon and $^{13}$C enriched inner
walls with 515 nm laser excitation. The G' mode of DWCNTs is
discussed in detail in Ref. \cite{PfeifferEPJB2004}: the upper G'
mode component corresponds to the outer tubes and the lower to the
inner tubes. The outer tube G' components are unaffected by the
$^{13}$C enrichment within the 1 cm$^{-1}$ experimental accuracy.
This gives an upper limit to the extra $^{13}$C in the outer wall of
1.4 \%. This proves that there is no sizeable carbon exchange
between the two walls as this would result in a measurable $^{13}$C
content on the outer wall, too.

The narrow RBMs of inner tubes and the freedom to control their
isotope enrichment allows to precisely compare the isotope related
phonon energy changes in the experiment and in \emph{ab-initio}
calculations. This was performed by J. K\"{u}rti and V. Z\'{o}lyomi
in Ref. \cite{SimonPRL2005}. The validity of the above simple
continuum model for the RBM frequencies was verified by performing
first principles calculations on the $(n,m)=(5,5)$ tube as an
example. In the calculation, the Hessian matrix was determined by
DFT using the Vienna Ab Initio Simulation Package
\cite{KressePRB1999}. Then, a large number of random $^{13}$C
distributions were generated and the RBM vibrational frequencies
were determined from the diagonalization of the dynamical matrix for
each individual distribution. It turns out that the calculation can
account for the above mentioned broadening of the RBM lines due to
the random distribution of the $^{12}$C and $^{13}$C nuclei
\cite{SimonPRL2005}.

The known characteristics of isotope labeled inner tubes allow to
study the possibility of inner tube growth from non-fullerene carbon
sources \cite{SimonCPL2006}. For this purpose, we chose organic
solvents containing aromatic rings, such as toluene and benzene.
These are known to wet the carbon nanotubes and are appropriate
solvents for fullerenes. As described in the following, the organic
solvents indeed contribute to the inner tube growth, however only in
the presence of C$_{60}$ ``stopper" molecules \cite{SimonCPL2006}.
In the absence of co-encapsulated fullerenes the solvents alone give
no inner tube.

The fullerene+organic solvents encapsulation was performed by
dissolving typically 150 $\mu $g fullerenes in 100 $\mu $l solvents
and then sonicating with 1 mg SWCNT in an Eppendorf tube for 1 h.
The weight uptake of the SWCNT is $\sim $15 \% \cite{SimonPRL2005}
that is shared between the solvent and the fullerenes. The peapod
material was separated from the solvent by centrifuging and it was
then greased on a sapphire substrate. The solvent prepared peapods
were treated in dynamic vacuum at 1250 $^{\circ}$C for 2 hours for
the inner tube growth. The inner tube growth efficiency was found
independent of the speed of warming.

\begin{figure}[tbp]
\begin{center}
\includegraphics[width=0.9\hsize]{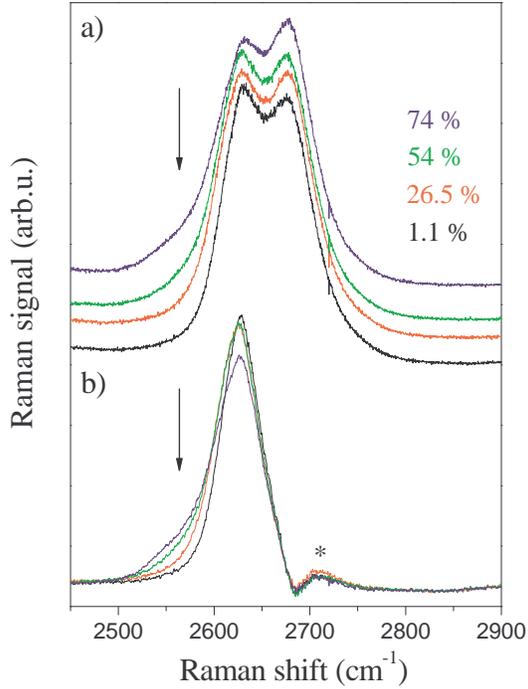}
\caption{a) The G' mode of toluene+C$_{60}$ peapod based DWCNTs with
varying $^{13}$C enrichment at 515 nm laser
excitation. From top to bottom: 74 \%, 54 \%, 26.5 \% and natural $%
^{13}$C content. b) The G' mode of the inner tubes after subtracting
the experimental SWCNT spectrum. A small residual peak is observed
around 2710 cm$^{-1}$ (denoted by an asterisk) due to the imperfect
subtraction. Arrows indicate the spectral weight shifted toward
lower frequencies. Reprinted figure with permission from Ref.
\cite{SimonCPL2006}, F. Simon and H. Kuzmany, Chem. Phys. Lett.
\textbf{425}, 85 (2006). Copyright (2006) by Elsevier.}
\label{CACReview_varying13C}
\end{center}
\end{figure}

The growth of inner tubes from the solvents can be best proven by
the use of C$_{60}$ containing natural carbon and a solvent mixture
consisting of $^{13}$C enriched and natural carbon containing
solvents with varying concentrations. Toluene was a mixture of ring
$^{13}$C labeled ($^{13}$C$_{6}$H$_{6}$-$^{\text{NAT}}$CH$_{3}$) and
natural toluene ($^{\text{NAT}}$C$_{7}$H$_{8}$). Benzene was a
mixture of $^{13}$C enriched and natural benzene. The labeled site
was $>$ 99 \% $^{13}$C labeled for both types of molecules. The
$^{13}$C content, $x$, of the solvent mixtures was calculated from
the concentration of the two types of solvents and by taking into
account the
presence of the naturally enriched methyl-group for the toluene. In Fig. \ref%
{CACReview_varying13C}a, we show the G' modes of DWCNTs with varying
$^{13}$C labeled content in toluene+C$_{60}$ based samples and in
Fig. \ref{CACReview_varying13C}b, we show the same spectra after
subtracting the outer SWCNT component. A shoulder appears for larger
values of $x$ on the low frequency side of the inner tube mode,
whereas the outer tube mode is unchanged. Similar behavior was
observed for the benzene+C$_{60}$ based peapod samples (spectra not
shown) although with a somewhat smaller spectral intensity of the
shoulder. The appearance of this low frequency shoulder is evidence
for the presence of a sizeable $^{13}$C content in the inner tubes.
This proves that the solvent indeed contributes to the inner tube
formation as it is the only sizeable source of $^{13}$C in the
current samples. The appearance of the low frequency shoulder rather
than the shift of the full mode indicates an inhomogeneous $^{13}$C
enrichment. A possible explanation is that smaller diameter
nanotubes might be higher $^{13}$C enriched as they retain the
solvent better than larger tubes.

\begin{figure}[tbp]
\begin{center}
\includegraphics[width=0.9\hsize]{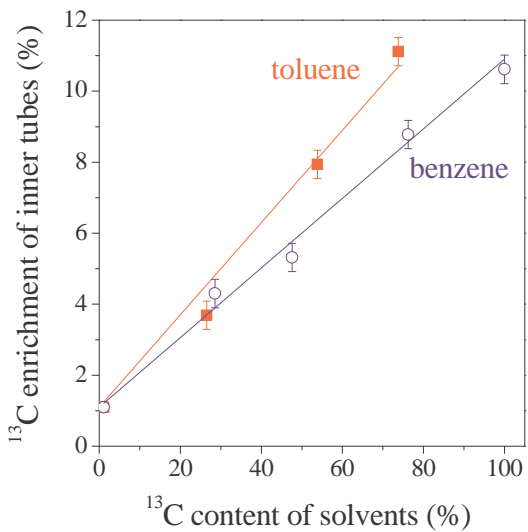}
\caption{$^{13}$C content of inner tubes based on the first moment
analysis as explained in the text as a function of $^{13}$C
enrichment of benzene and toluene. Lines are linear fits to the data
and are explained in the text. Reprinted figure with permission from
Ref. \cite{SimonCPL2006}, F. Simon and H. Kuzmany, Chem. Phys. Lett.
\textbf{425}, 85 (2006). Copyright (2006) by Elsevier.}
\label{CACReview_13Cscaling}
\end{center}
\end{figure}

To quantify the $^{13}$C enrichment of the inner tubes, the
downshifted spectral weight of the inner tube G' mode was determined
from the subtracted spectra in Fig. \ref{CACReview_13Cscaling}b. The
subtraction does not give a flat background above 2685 cm$^{-1}$,
however it is the same for all samples and has a small spectral
weight, thus it does not affect the current analysis. The
line-shapes strongly deviate from an ideal Lorentzian profile.
Therefore the line positions cannot be determined by fitting,
whereas the first moments are well defined quantities. The effective
$^{13}$C enrichment of the inner tubes, $c$, is calculated from
$\left( \nu _{0}-\nu \right) /\nu
_{0}=1-\sqrt{\frac{12+c_{0}}{12+c}}$, where $\nu _{0}$ and $\nu $
are the first moments of the inner tube G' mode in the natural
carbon and enriched materials, respectively, and $c_{0}=0.011$ is
the natural abundance of $^{13} $C in carbon. The validity of this
``text-book formula" is discussed above and it was verified by
\emph{ab-initio} calculations for enriched inner tubes in Ref.
\cite{SimonPRL2005}. In Fig. \ref{CACReview_13Cscaling}, we show the
effective $^{13}$C content in the inner tubes as a function of the
$^{13}$C content in the starting solvents. The scaling of the
$^{13}$C content of the inner tubes with that in the starting
solvents proves that the source of the $^{13}$C is indeed the
solvents. The highest value of the relative shift for the toluene
based material, $\left( \nu _{0}-\nu \right) /\nu _{0}=0.0041(2)$,
corresponds to about 11 cm$^{-1}$ shift in the first moment of the
inner
tube mode. The shift in the radial breathing mode range (around 300 cm$^{-1}$%
) \cite{DresselhausTubesNew} would be only 1 cm$^{-1}$. This
underlines why the
high energy G' mode is convenient for the observation of the moderate $^{13}$%
C enrichment of the inner tubes. When fit with a linear curve with
$c_{0}+A\ast x$, the slope, $A$ directly measures the carbon
fraction in the inner tubes that originates from the solvents.

The synthesis of inner tubes from organic solvent proves that any
form of carbon that is encapsulated inside SWCNTs contributes to the
growth of inner tubes. As mentioned above, inner tubes are not
formed in the absence of fullerenes but whether the fullerene is
C$_{60}$ or C$_{70}$ does not play a role. It suggests that
fullerenes act only as a stopper to prevent the solvent from
evaporating before the synthesis of the inner tube takes place. It
also clarifies that the geometry of fullerenes do not play a
distinguished role in the inner tube synthesis as it was originally
suggested \cite{SmalleyPRL2002,TomanekPRB}. It also proves that
inner tube growth can be achieved irrespective of the carbon source,
which opens a new prospective to explore the in-the-tube chemistry
with other organic materials.

\subsection{ESR studies on encapsulated magnetic fullerenes}

Observation of the intrinsic ESR signal of pristine SWCNTs remains
elusive \cite{NemesPRB2000,SalvetatPRB2005}. Now, it is generally
believed that intrinsic ESR of the tubes can not be observed as
conduction electrons on metallic tubes are relaxed by defects too
fast to be observable. In addition, one always observes a number of
ESR active species in a sample, such as graphitic carbon or magnetic
catalyst particles, which prevent a meaningful analysis of the
signal. In contrast, local probe studies could still allow an ESR
study of tubes, provided the local spin probe can be selectively
attached to the tubes. This goal can be achieved by using magnetic
fullerenes, such as e.g. N@C$_{60}$ or C$_{59}$N, since fullerenes
are known to be selectively encapsulated inside SWCNTs
\cite{SmithNAT} and can be washed from the outside using organic
solvents \cite{KatauraSM2001}. As the properties and handling of the
two magnetic fullerenes are quite different, the synthesis of the
corresponding peapods and the results are discussed separately.

N@C$_{60}$ is an air stable fullerene \cite{WeidingerPRL} but decays
rapidly above $\sim$ 200 $^{\circ}$C \cite{WaiblingerPRB} which
prevents the use of the conventional vapor method of peapod
preparation which requires temperatures above 400 $^{\circ}$C. To
overcome this limitation and to allow in general the synthesis of
temperature sensitive peapod materials, low temperature peapod
synthesis (solvent method) was developed independently by four
groups \cite{YudasakaCPL,SimonCPL2004,Monthioux2004,BriggsJMC}.
These methods share the common idea of mixing the opened SWCNTs with
C$_{60}$ in a solvent with low fullerene solubility such as methanol
\cite{YudasakaCPL} or n-pentane \cite{SimonCPL2004}. The
encapsulation is efficient as it is energetically preferred for
C$_{60}$ to enter the tubes rather than staying in the solution.
After the solvent filling, excess fullerenes can be removed by
sonication in toluene, which is a good fullerene solvent as it was
found that fullerenes enter the tube irreversibly. HR-TEM has shown
an abundant filling with the fullerenes
\cite{YudasakaCPL,Monthioux2004} and a more macroscopic
characterization using Raman spectroscopy has proven that peapods
prepared by the solvent method are equivalent to the vapor prepared
peapods \cite{SimonCPL2004}.

\begin{figure}[tbp]
\begin{center}
\includegraphics[width=0.9\hsize]{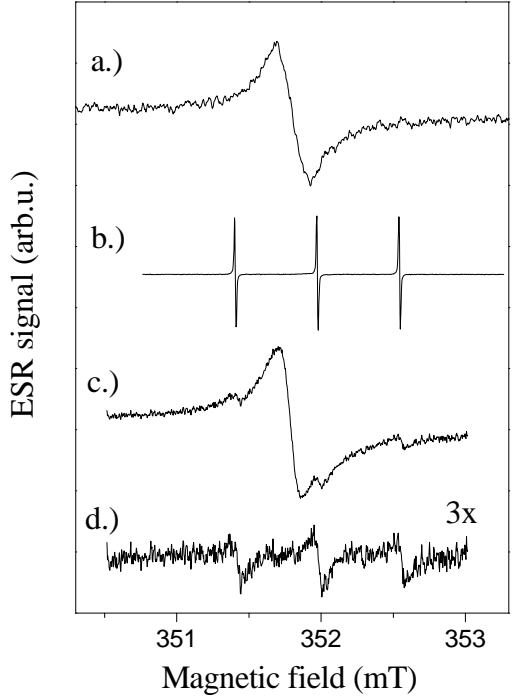}
\caption{X-band electron spin resonance spectrum of the a.) pristine
SWCNT, b.) crystalline N@C$_{60}$:C$_{60}$, c.)
(N@C$_{60}$:C$_{60}$)@SWCNT and d.) the triplet component of the
(N@C$_{60}$:C$_{60}$)@SWCNT ESR spectrum at ambient temperature.
Reprinted figure with permission from Ref. \cite{SimonCPL2004}, F.
Simon \textit{et al.} Chem. Phys. Lett. \textbf{383}, 362 (2004).
Copyright (2004) by Elsevier.} \label{CACReview_lowTfillesr}
\end{center}
\end{figure}

The low temperature synthesis allows to encapsulate the N@C$_{60}$
fullerene. The N@C$_{60}$:C$_{60}$ endohedral fullerene: fullerene
solid solution can be produced in a N$_{2}$ arc-discharge tube
following Ref. \cite{PietzakCPL} with a typical yield of a few 10
ppm \cite{JanossyKirch2000}. In Fig. \ref{CACReview_lowTfillesr}.,
the ESR spectra of the starting SWCNT, (N@C$_{60}$:C$_{60}$)@SWCNT,
and N@C$_{60}$:C$_{60}$ are shown. The ESR spectrum of the pristine
SWCNT for the magnetic field range shown is dominated by a signal
that is assigned to some residual carbonaceous material, probably
graphite. Fig. \ref{CACReview_lowTfillesr}c. shows, that after the
solvent encapsulation of N@C$_{60}$:C$_{60}$ in the NCL-SWCNT, a
hyperfine N triplet ESR is observed, similar to that in pristine
N@C$_{60}$:C$_{60}$, superimposed on the broad signal present in the
pristine nanotube material. Fig. \ref{CACReview_lowTfillesr}d. shows
the triplet component of this signal after subtracting the signal
observed in pristine SWCNT. The hyperfine triplet in
N@C$_{60}$:C$_{60}$ is the result of the overlap of the
$^{4}$S$_{3/2}$ state of the three 2p electrons of the N atom and
the $^{14}$N nucleus, with nuclear spin, $I=1$. The isotropic
hyperfine coupling of N@C$_{60}$:C$_{60}$ is unusually high as a
result of the strongly compressed N atomic 2p$^{3}$ orbitals in the
C$_{60}$ cage thus it unambiguously identifies this material
\cite{WeidingerPRL}. The hyperfine coupling constant observed for
the triplet structure in the encapsulated material,
$A_{\text{iso}}=0.57\pm 0.01 $ mT, agrees within experimental
precision with that observed in N@C$_{60}$:C$_{60}$
\cite{WeidingerPRL}, which proves that the encapsulated material is
(N@C$_{60}$:C$_{60}$)@SWCNT. The ESR line-width for the encapsulated
material, $\Delta H_{pp}$ = 0.07 mT, is significantly larger than
the resolution limited $\Delta H_{pp}$ =0.01 mT in the pristine
N@C$_{60}$:C$_{60}$ material, the lines being Lorentzian. The most
probable cause for the broadening is static magnetic fields from
residual magnetic impurities in the SWCNT \cite{TangNMRSCI}. The ESR
signal intensity is proportional to the number of N spins, and this
allows the quantitative comparison of N concentrations in
(N@C$_{60}$:C$_{60}$)@SWCNT and N@C$_{60}$:C$_{60}$. It was found
that the number of observed N@C$_{60}$ spins is consistent with the
number expected from a good filling efficiency \cite{SimonCPL2004}.

As seen from the ESR results on encapsulated N@C$_{60}$, relatively
limited information can be deduced about the tubes themselves. This
stems from the fact that the N spins are well shielded in N@C$_{60}$
\cite{DinsePCCP} and are thus are relatively insensitive to the
SWCNT properties. In contrast, N@C$_{60}$@SWCNT peapods might find
another application as building elements of a quantum computer as
proposed by Harneit \textit{et al.} \cite{HarneitPSS}. A better
candidate for magnetic fullerene encapsulation is the C$_{59}$N
monomer radical as here the unpaired electron is on the cage and is
a sensitive probe of the environment. This material can be
chemically prepared \cite{WudlSCI}, however it forms a non-magnetic
dimer crystal (C$_{59}$N)$_{2}$. It appears as a spinless monomer in
an adduct form \cite{WudlReview} or attached to surface dangling
bonds \cite{PrassidesPRL}. The magnetic C$_{59}$N monomer
radical is air sensitive but it can be stabilized as a radical when it is dilutely mixed in C$_{60}$\cite%
{FulopCPL}. As a result, a different strategy has to be followed to
encapsulate C$_{59}$N inside SWCNT, which is discussed in the
following along with preliminary ESR results \cite{SimonCAR2006}.

To obtain C$_{59}$N peapods, air stable C$_{59}$N derivatives,
(C$_{59}$N-der in the following) were prepared chemically by A.
Hirsch and F. Hauke following standard synthesis routes
\cite{WudlReview,HirschC59NReview}. The C$_{59}$N-der was
4-Hydroxy-3,5-dimethyl-phenyl-hydroazafullerene. The C$_{59}$N
derivatives were encapsulated either pure or mixed with C$_{60}$ as
C$_{59}$N-der:C$_{60}$ with 1:9 concentrations using a modified
version of the low temperature encapsulation method. In brief, the
mixture of the dissolved fullerenes and SWCNTs were sonicated in
toluene and filtered. It is expected that the C$_{59}$N monomer
radical can be obtained after a heat treatment in dynamic vacuum,
which is discussed below.

Raman spectroscopy was performed to characterize the SWCNT filling
with the C$_{59}$N-der \cite{SimonCAR2006}. The major Raman modes of the pristine C$%
_{59}$N-der are similar to those of the (C$_{59}$N)$_{2}$ dimer \cite%
{KuzmanyPRB1999}. The strongest mode is observed at 1459.2 cm$^{-1}$
which is derived from the C$_{60}$ $A_{\text{g}}$(2) mode and is
downshifted to 1457 cm$^{-1}$ after the encapsulation procedure. The
2.2 cm$^{-1}$ downshift proves the encapsulation of the molecule
inside the SWCNT. When encapsulated inside SWCNTs, the corresponding
A$_{\text{g}}$(2) mode of C$_{60}$ downshifts with 3 cm$^{-1}$ ,
which is assigned to the softening of the C$_{60}$ $A_{\text{g}}$(2)
vibrational mode due to the interaction between the ball and the
SWCNT wall \cite{PichlerPRL2001}.

\begin{figure}[tbp]
\begin{center}
\includegraphics[width=0.9\hsize]{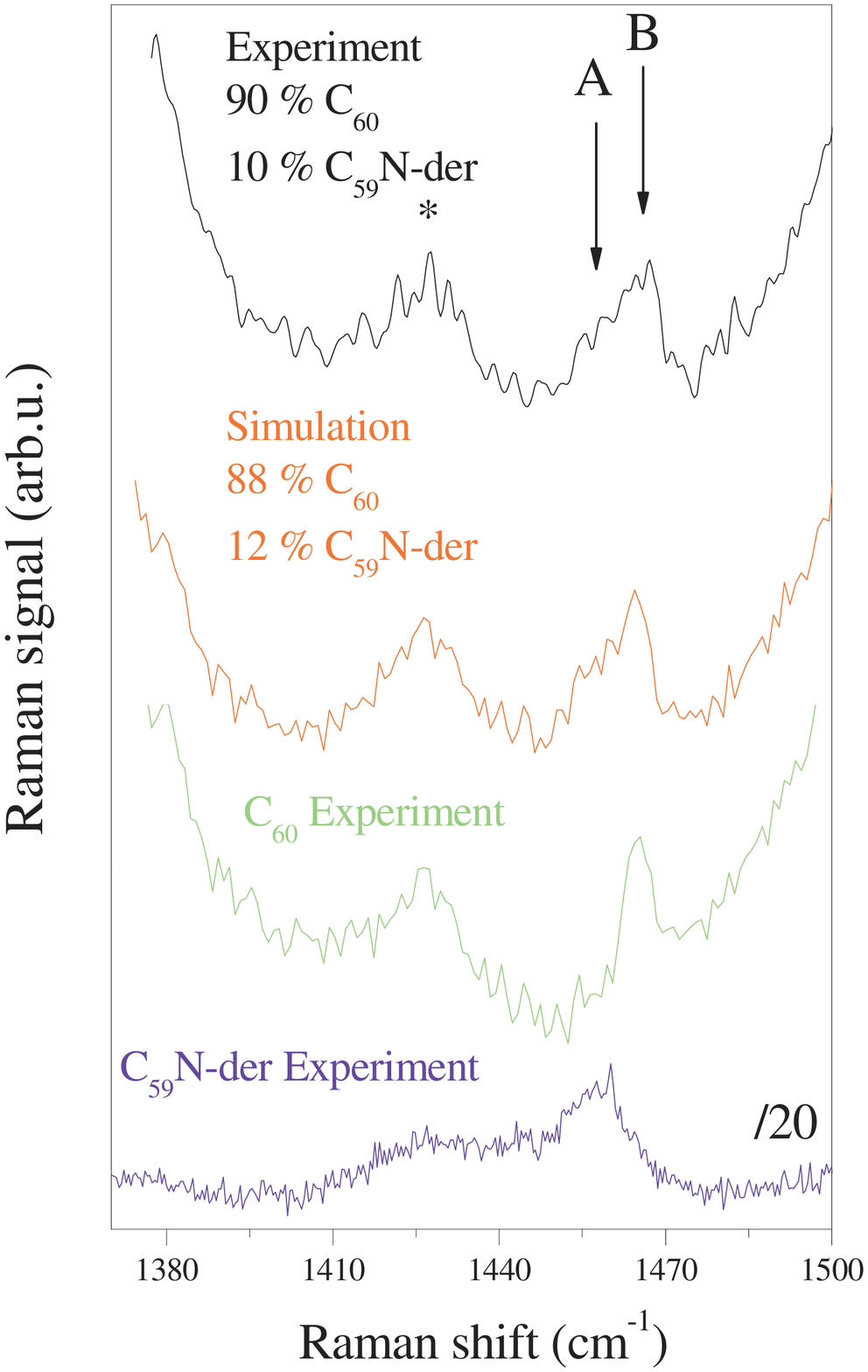}
\caption{Raman spectra of the encapsulated C$_{59}$N-der:C$_{60}$
mixture at the 488 nm laser excitation. The spectra for the
C$_{59}$N-der and C$_{60}$ peapods is shown together with their
weighted sum as explained in the text. A and B mark the components
coming nominally from the superposing two phases. The asterisk marks
a mode that is present in the pristine SWCNT material. Note the
different scale for the C$_{59}$N-der peapod material. Reprinted
figure with permission from Ref. \cite{SimonCAR2006}, F. Simon
\textit{et al.} Carbon, \textbf{44}, 1958 (2006). Copyright (2006)
by Elsevier.} \label{CACReview_C59Ndermixedpeapodspectra}
\end{center}
\end{figure}

The integrated intensity of the observed A$_{\text{g}}$(2) derived mode of the C$%
_{59}$N is approximately 5 times larger than that of a C$_{60}$
peapod prepared identically when normalized by the SWCNT G mode
intensity. This, however, can not be used to measure the
encapsulation efficiency as Raman intensities depend on the strength
of the Raman resonance enhancement and the Raman scattering matrix
elements \cite{KuzmanyBook}. For C$_{60}$ peapods the Raman signal
was calibrated with independent and carbon number sensitive
measurements: EELS studies gave the total number of C$_{60}$ related
and non-C$_{60}$ related carbons \cite{LiuPRB2002} and the mass of
encapsulated C$_{60}$s was determined from NMR studies using
$^{13}$C enriched fullerenes \cite{SimonPRL2005,SingerPRL2005}. In
the current case, neither methods can be employed and we determined
the filling efficiency for the azafullerene by encapsulating a
mixture of the azafullerene and C$_{60}$. In Fig.
\ref{CACReview_C59Ndermixedpeapodspectra}, the Raman spectra of the
encapsulated C$_{59}$N-der:C$_{60}$ mixture with weight ratios of
1:9 in the starting solvent is shown. The Raman spectrum of the
encapsulated mixture was simulated with a weighted sum of the
separately recorded spectra for encapsulated C$_{59}$N-der and
C$_{60}$. The best agreement between the simulated and the
experimental spectra is for a C$_{59}$N-der content of 0.12(2). This
value is close to the expected value of 0.1 and it proves that the
azafullerene enters the tubes with the same efficiency as C$_{60}$.

\begin{figure}[tbp]
\begin{center}
\includegraphics[width=0.9\hsize]{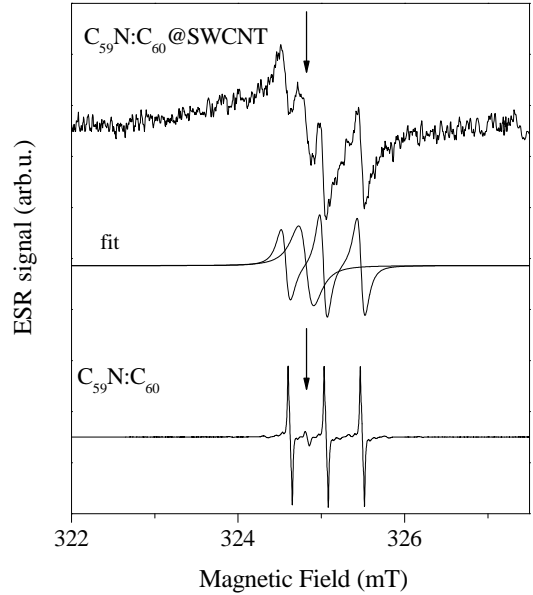}
\caption{ESR spectra of crystalline C$_{59}$N:C$_{60}$ (a) and (C$_{59}$N:C$_{60}$%
)@SWCNT obtained by annealing the (C$_{59}$N-der:C$_{60}$)@SWCNT.
Solid curves show the deconvolution of the different ESR components
for the encapsulated material. Reprinted figure with permission from
Ref. \cite{SimonPRL2005}, F. Simon \textit{et al.} Phys. Rev. Lett.
\textbf{97}, 136801 (2006). Copyright (2006) by the American
Physical Society.} \label{CACReview_C59N_ESR}
\end{center}
\end{figure}

Fig. \ref{CACReview_C59N_ESR} shows the room temperature ESR spectra
of C$_{59}$N:C$_{60}$@SWCNT after 600 $^{\circ }$C vacuum annealing
from Ref. \cite{SimonPRL2006}. The spectra of C$_{59}$N:C$_{60}$, a
C$_{59}$N monomer embedded in C$_{60}$ \cite{FulopCPL}, is also
shown for comparison. This latter spectrum was previously assigned
to the superposition of rotating C$_{59}$N monomers and bound
C$_{59}$N-C$_{60}$ heterodimers \cite{RockenbauerPRL2005}. The large
spin density at the $^{14}$N nucleus of the rotating C$_{59}$N
molecule results in an ESR triplet signal and the C$_{59}$N-C$_{60}$
heterodimer has a singlet signal (arrow in Fig.
\ref{CACReview_C59N_ESR}) as the spin density resides on the
C$_{60}$ molecule. $^{14}$N triplet structures are observed in the
peapod samples with identical hyperfine coupling as in the
crystalline sample and are thus identified as the ESR signals of
rotating C$_{59}$N monomer radicals encapsulated inside SWCNTs. The
additional component (arrow in Fig. \ref{CACReview_C59N_ESR})
observed for
sample B, which contains co-encapsulated C$_{60}$, is identified as C$_{59}$%
N-C$_{60}$ heterodimers encapsulated inside SWCNTs since this signal
has the same $g$-factor as in the crystalline material. This singlet
line is absent in sample A which does not contain C$_{60}$. For both
peapod samples a broader line with HWHM of $\Delta H \sim 0.6$ mT is
also observed. The broader component appears also on heat treatment
of reference samples without encapsulated C$_{59}$N-der and is
identified as a side-product. Annealing at 600 $^{\circ }$C is
optimal: lower temperatures result in smaller C$_{59}$N signals and
higher temperatures increase the broad impurity signal without
increasing the C$_{59}$N intensity.

The observation of the ESR signal of C$_{59}$N related spins proves
that after the 600 $^{\circ }$C heat treatment, a sizeable amount of
rotating C$_{59}$N monomer radicals are present in the sample. This
is not surprising in the view of the ability to form C$_{59}$N
monomers from C$_{59}$N at similar temperatures \cite{SimonJCP},
however the current process is not reversible and the remnants of
the side-groups are probably removed by the dynamic pumping.

\subsection{NMR studies on isotope engineered heteronuclear nanotubes}

The growth of the ``isotope engineered" nanotubes, i.e. DWCNTs with
highly enriched inner wall allows to study the electronic properties
of small diameter carbon nanotubes with an unprecedented specificity
using NMR. For normal SWCNTs, either grown from natural or $^{13}$C
enriched carbon, the NMR signal originates from all kinds of carbon
like amorphous or graphitic carbon.

\begin{figure}[tbp]
\begin{center}
\includegraphics[width=0.9\hsize]{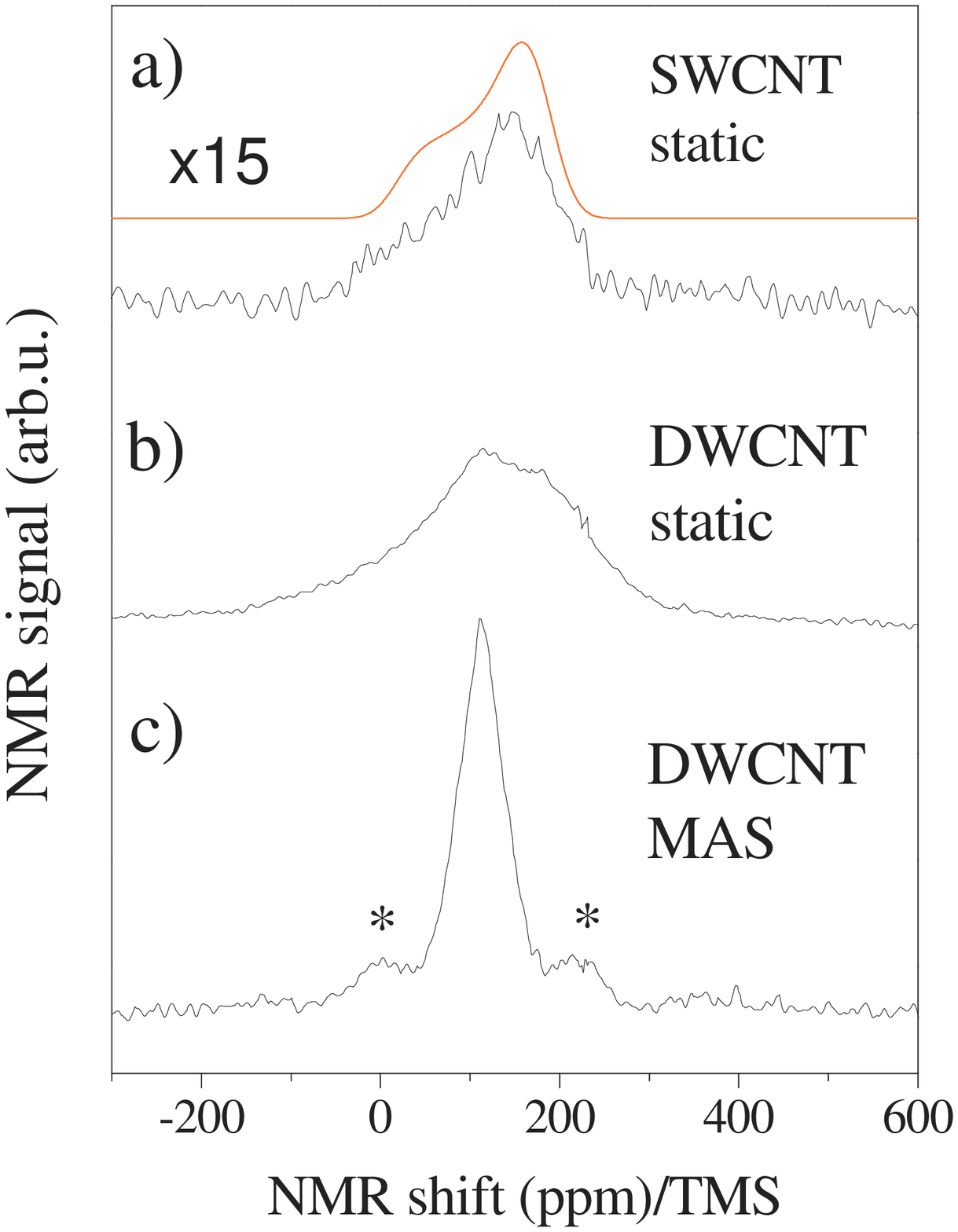}
\caption{NMR spectra normalized by the total sample mass, taken with
respect to the tetramethylsilane (TMS) shift. (a) Static spectrum
for non-enriched SWCNT enlarged by 15. Smooth solid line is a
chemical shift anisotropy powder pattern
simulation with parameters published in the literature \protect\cite%
{TangNMRSCI}. (b) Static and (c) MAS spectra of
$^{13}$C$_{0.89}$-DWCNT, respectively. Asterisks show the sidebands
at the 8 kHz spinning frequency. Reprinted figure with permission
from Ref. \cite{SimonPRL2005}, F. Simon \textit{et al.} Phys. Rev.
Lett. \textbf{95}, 017401 (2005). Copyright (2005) by the American
Physical Society.} \label{CACReview_Static_NMR}
\end{center}
\end{figure}

NMR allows to determine the macroscopic amount of enriched tubes as
it is sensitive to the number of $^{13}$C nuclei in the sample. In
Fig. \ref{CACReview_Static_NMR}, we show the static and magic angle
spinning spectra of $^{13}$C enriched DWCNTs, and the static
spectrum for the SWCNT material. The mass fraction which belongs to
the highly enriched phase can be calculated from the integrated
signal intensity by comparing it to the signal intensity of the 89
\% $^{13}$C enriched fullerene material. It was found that the mass
fraction of the highly enriched phase relative to the total sample
mass is 13(4) \% which agrees with the expected value of 15 \%. The
latter is obtained from the SWCNT purity (50 \%), $\sim $70 \%
volume filling for peapod samples \cite{LiuPRB2002}, and the mass
ratio of encapsulated fullerenes to the mass of the SWCNTs. This
suggests that the NMR signal comes nominally from the inner tubes,
and other carbon phases such as amorphous or graphitic carbon are
non $^{13}$C enriched.

The typical chemical shift anisotropy (CSA) powder pattern was
observed for
the SWCNT sample in agreement with previous reports \cite%
{TangNMRSCI,GozeBacCAR2002}. However, the static DWCNT spectrum
cannot be explained with a simple CSA powder pattern even if the
spectrum is dominated by the inner tube signal. The complicated
structure of the spectrum suggests that the chemical shift tensor
parameters are highly distributed for the inner tubes. It is the
result of the higher curvature of inner tubes as compared to the
outer ones:\ the variance of the diameter distribution is the same
for the inner and outer tubes \cite{SimonDWCNTReview} but the
corresponding bonding angles show a larger variation
\cite{KuertiNJP2003}. In addition, the residual line-width in the
MAS experiment, which is a measure of the sample inhomogeneity, is
60(3) ppm, i.e. about twice as large as the $\sim $35 ppm found
previously for SWCNT samples \cite{TangNMRSCI,GozeBacCAR2002}. The
isotropic line position, determined from the MAS measurement, is
111$(2)$ ppm. This value is significantly smaller than the isotropic
shift of the SWCNT samples of 125 ppm
\cite{TangNMRSCI,GozeBacCAR2002}. However, recent theoretical
\emph{ab-initio} calculations by F. Mauri and co-workers have
successfully explained this anomalous isotropic chemical shift
\cite{MauriPRB2006}. It was found that diamagnetic demagnetizing
currents on the outer walls cause the diamagnetic shift of the inner
tube NMR signal.

In addition to the line position, dynamics of the nuclear relaxation
is a sensitive probe of the local electronic properties
\cite{SlichterBook}. The electronic properties of the nanotubes was
probed using the spin lattice relaxation time, $T_{1}$, defined as
the characteristic time it takes the $^{13}$C nuclear magnetization
to recover after saturation \cite{SingerPRL2005}. The signal
intensity after saturation, $S(t)$, was deduced by integrating the
fast Fourier transform of half the spin-echo for different delay
times $t$. The data were taken with excitation pulse lengths $\pi/2
=$ 3.0 $\mu$s and short pulse separation times of $\tau=$ 15 $\mu$s
\cite{SlichterBook}. The value of $T_{1}$ was obtained by fitting
the $ t$ dependence of $S(t)$ to the form $ S(t) \ = \ S_{a} - S_{b}
\cdot M(t)$, where $S_{a} \simeq S_{b}$ ($>0$) are arbitrary signal
amplitudes, and
\begin{equation}
M(t) = \exp\left[-\left(t/T_{1}^{e}\right)^{\beta}\right],
\label{CACReview_stretched}
\end{equation}
is the reduced magnetization recovery of the $^{13}$C nuclear spins.
It was found that $M(t)$ does not follow the single exponential form
with $\beta=1$, but instead fits well to a stretched exponential
form with $\beta \simeq 0.65(5)$, implying a distribution in the
relaxation times $T_{1}$. For a broad range of experimental
conditions, the upper 90 \% of the $M(t)$ data is consistent with
constant $\beta \simeq 0.65(5)$, implying a field and temperature
independent underlying distribution in $T_{1}$. The collapse of the
data with constant $\beta = 0.65(5)$ is a remarkable experimental
observation and it implies that each inner-tube in the powder sample
has a different value of $T_{1}$, yet \emph{all} the $T_{1}$
components and therefore all the inner-tubes follow the same $T$ and
$H$ dependence within experimental uncertainty. This finding is in
contrast to earlier reports in SWCNTs where $M(t)$ fits well to a
bi-exponential distribution, 1/3 of which had a short $T_{1}$ value
characteristic of fast relaxation from metallic tubes, and the
remaining 2/3 had long $T_{1}$ corresponding to the semiconducting
tubes
\cite{TangNMRSCI,GozeBacCAR2002,ShimodaPRL2002,KleinhammesPRB2003},
as expected from a macroscopic sample of SWCNTs with random
chiralities.

\begin{figure}
\begin{center}
\includegraphics[width=1\hsize]{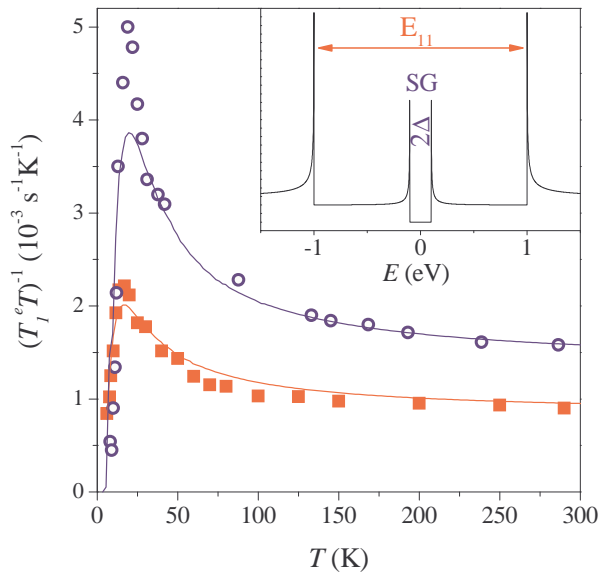}
\caption{Temperature dependence of spin-lattice relaxation rate
divided by temperature, $ 1/T_{1}^{e}T$, in units of $(10^{3}\times$
s$^{-1}$ K $^{-1}$) for 3.6 T ($\bigcirc$) and 9.3 T
($\blacksquare$). Grey curves are best fits to Eq.
(\ref{CACReview_gapT1}) with $2\Delta = 46.8 (40.2)$ K for $H= 3.6
(9.3)$ Tesla, respectively. Inset shows the suggested DOS with a
small energy, $2\Delta$ secondary gap at the Fermi level of metallic
inner tubes, which is displayed not to scale. Note the van Hove
singularities (vHs) at $\pm\Delta$. Reprinted figure with permission
from Ref. \cite{SingerPRL2005}, P. M. Singer \textit{et al.} Phys.
Rev. Lett. \textbf{95}, 236403 (2005). Copyright (2005) by the
American Physical Society.} \label{CACReview_T1T_DOS}
\end{center}
\end{figure}

The bulk average $T_{1}^{e}$ defined in Eq.
(\ref{CACReview_stretched}) can be considered and its uniform $T$
and $H$ dependence can be followed. The $M(t)$ data can be fitted
with the constant exponent $\beta=0.65(5)$, which reduces
unnecessary experimental scattering in $T_{1}^{e}$. In Fig.
\ref{CACReview_T1T_DOS}, we show the temperature dependence of
$1/T_{1}^{e}T$ for two different values of the external magnetic
field, $H$. The data can be separated into two temperature regimes;
the high temperature regime $\gtrsim$ 150 K, and the low $T$ regime
$\lesssim$ 150 K. At high temperatures, $1/T_{1}^{e}T$ is
independent of $T$ which indicates a metallic state
\cite{SlichterBook} for all of the inner tubes. A strong magnetic
field dependence for $T_{1}$ was also observed, which was explained
by a 1D spin diffusion mechanism for $T_{1}$ \cite{SingerPRL2005}.

The experimentally observed uniform metallicity of inner tubes is a
surprising observation. This was suggested to be caused by the
shifting of the inner tube Fermi levels due to charge transfer
between the two tube walls. Indeed, using \textit{ab-initio}
calculations Okada and Oshiyama have found that DWCNTs made of
non-metallic zig-zag inner-outer tubes, such as the (7,0)@(16,0)
DWCNT, are metallic \cite{OkadaPRL2003}. The direction of the charge
transfer goes against the Faraday effect as inner tubes are electron
and outer tubes are hole doped. Although, calculations are difficult
if not impossible for an arbitrary inner-outer tube pair, this
result confirms that two non-metallic tubes when producing an inner
tube can render the electronic structure metallic.

The origin of the unusual $T$ dependence of 1/$T_{1}^{e}T$ in the
low temperature regime ($\lesssim 150$ K) is peculiar. Some
explanations can be ruled out as its origin. Firstly, one can rule
out the possibility of an activation type mechanism where $T_{1}$ is
dominated by fluctuating hyperfine fields with a characteristic time
scale $\tau $ which increases with decreasing $T$ (i.e. glassy
slowing). This would result in a peaked relaxation with a strongly
field dependent peak value \cite{SlichterBook}, which is clearly not
the case. Furthermore at low $T$, 1/$T_{1}^{e}T$ drops below its
high temperature value, which rules out the possibility of a $T $
independent component in 1/$T_{1}^{e}T$ plus an activated component
on top. Secondly, the possibility of a simple $1/T $ Curie-like $T$
dependence in 1/$T_{1}^{e}T$ as a result of paramagnetic centers in
the sample can be ruled out. This can be inferred from the
pronounced gap in 1/$ T_{1}^{e}T$, together with the fact that no
loss of $^{13}$C NMR signal intensity in the entire temperature
range of the experiment was observed.

The simplest possible explanation for the experimental data is a
non-interacting electron model of a 1D semiconductor with a small
secondary gap (SG). The 1/$ T_{1}^{e}T$ data can be fitted using
this model with only one parameter, the homogeneous SG, 2$\Delta $.
The normalized form of the gapped 1D density-of-states $n(E)$
\begin{equation}
n(E)\ =\
\begin{cases}
\frac{E}{\sqrt{E^{2}-\Delta ^{2}}} & \text{for }|E|>\Delta \cr0 &
\text{ otherwise}\cr
\end{cases}
\label{CACReview_newDOS}
\end{equation}
\noindent here, $E$ is taken with respect to the Fermi energy). The
total DOS of an inner tube is shown schematically in the insert of
Fig. \ref{CACReview_T1T_DOS}. Eq. (\ref{CACReview_newDOS}) is used
to calculate 1/$T_{1}^{e}T$ \cite{Moriya} as such
\begin{equation}
\frac{1}{T_{1}^{e}T}\ =\ \alpha (\omega )\int_{-\infty }^{\infty
}n(E)n(E+\omega )\left( -\frac{\delta f}{\delta E}\right) dE,
\label{CACReview_gapT1}
\end{equation}
where $E$ and $\omega$ are in temperature units for clarity, $f$ is
the Fermi function $f = [\exp(E/T) + 1 ]^{-1}$, and the amplitude
factor $\alpha(\omega)$ is the high temperature value for
$1/T_{1}^{e}T$. The results of the best fit of the data to Eq.
(\ref{CACReview_gapT1}) are presented in Fig.
\ref{CACReview_T1T_DOS}, where $2\Delta = 43(3)$ K ($\equiv$ 3.7
meV) is $H$ independent within experimental scattering between 9.3
and 3.6 Tesla.

The origin of the experimentally observed gap still remains to be
clarified. Tight binding calculations predict that applied magnetic
fields can induce SG's of similar magnitude for metallic SWCNT
\cite{LuPRL1995}. However, such a scenario can be excluded due to
the absence of any field dependence of the gap. The NMR data would
be more consistent with a curvature induced SG for metallic tubes
\cite {HamadaPRL1992,KanePRL1997,MintmirePRL1998,ZolyomiPRB2004},
however for the typical inner-tubes the predicted values, $\sim 100$
meV, are over an order of magnitude larger than our experimental
data. Other scenarios, such as quantization of levels due to finite
short lengths of the nanotubes could be considered as well, however,
in all these cases a behavior independent of tube size and chirality
is certainly not expected.

This suggests that electron-electron interactions may play an
important role for the metallic inner tubes. It has been predicted
that electron-electron correlations and a Tomonaga-Luttinger (TLL)
state leads to an increase in 1/$T_{1}T$ with decreasing $T$
\cite{Yoshioka}, which is a direct consequence of the 1D electronic
state. The correlated 1D nature may also lead to a Peierls
instability \cite{DresselhausTubes} with the opening of a small
collective gap $2 \Delta$ and a sharp drop in 1/$T_{1}T$ below
$\Delta \sim 20$ K. Therefore, the presence of both a TLL state and
a Peierls instability could possibly account for the data.

Summarizing the NMR studies, it was shown that $T_{1}$ has a similar
$T$ and $H$ dependence for all the inner-tubes, with no indication
of a metallic/semiconducting separation due to chirality
distribution. At high temperatures, ($T\gtrsim 150$ K)
1/$T_{1}^{e}T$ of the inner tubes exhibit a metallic 1D spin
diffusion state. Below $\sim $150 K, 1/$T_{1}^{e}T$ increases
dramatically with decreasing $T$, and a gap in the spin excitation
spectrum is found below $ \Delta \simeq $ 20 K, which is suggested
to be caused by a Peierls instability \cite
{DresselhausTubes,DresselhausTubesNew}.

\section{Summary}
In summary, we reviewed how in-the-tube functionalization of SWCNTs
can be used to study various properties of the tubes themselves.
Inner tubes grown from encapsulated fullerenes were shown to be an
excellent probe of diameter dependent reactions on the outer tubes.
Inner tubes grown from isotope labeled fullerenes and organic
solvents allowed to understand the role of the different carbon
phases in the growth of the inner tubes. In addition, isotope
labeled inner tubes were shown to yield an unparalleled precision to
study the density of states near the Fermi level using NMR. It was
reviewed how magnetic fullerenes can be encapsulated inside SWCNTs
yielding linear spin chains with sizeable spin concentrations and
also to allow ESR studies of the tube properties.

\section{Acknowledgements}
This work was supported by the Austrian Science Funds (FWF) project
Nr. 17345, by the Hungarian State Grants No. TS049881, F61733 and
NK60984, by the EU projects BIN2-2001-00580 and MERG-CT-2005-022103,
by the Zolt\'{a}n Magyary and Bolyai postdoctoral fellowships. J.
Bernardi and Ch. Schaman are acknowledged for the HR-TEM
contribution and for preparing some of the figures, respectively.

\dag Present address: Budapest University of Technology and
Economics, Institute of Physics and Solids in Magnetic Fields
Research Group of the Hungarian Academy of Sciences, H-1521,
Budapest P.O.Box 91, Hungary

\bibliographystyle{unsrt}

\bibliography{CACReview}

\begin{thebibliography}{100}

\bibitem{IijimaNAT1991}
Sumio Iijima.
\newblock {Helical microtubules of graphitic carbon}.
\newblock {\em Nature}, 354:56--58, 1991.

\bibitem{IijimaNAT1993}
Sumio Iijima and Toshinari Ichihashi.
\newblock {Single-shell carbon nanotubes of 1-nm diameter}.
\newblock {\em Nature}, 363:603--605, 1993.

\bibitem{BethuneNAT1993}
D.~S. Bethune, C.~H. Kiang, M.~S. DeVries, G.~Gorman, Savoy R., and R.~Beyers.
\newblock {Cobalt-catalysed growth of carbon nanotubes with single-atomic-layer
  walls}.
\newblock {\em Nature}, 363:605, 1993.

\bibitem{Obraztsov}
A.~N. Obraztsov, I.~Pavlovsky, A.~P. Volkov, E.~D. Obraztsova, A.~L. Chuvilin,
  and V.~L. Kuznetsov.
\newblock {}.
\newblock {\em J. Vac. Sci. Techn. B}, 18:1059, 2000.

\bibitem{ZhouAPL2002}
G.~Z. Yue, Q.~Qiu, B.~Gao, Y.~Cheng, J.~Zhang, H.~Shimoda, S.~Chang, J.~P. Lu,
  and O.~Zhou.
\newblock {Generation of continuous and pulsed diagnostic imaging x-ray
  radiation using a carbon-nanotube-based field-emission cathode}.
\newblock {\em Appl. Phys. Lett.}, 81:355--368, 2002.

\bibitem{HafnerNAT}
J.~H. Hafner, C.~L. Cheung, and C.~M. Lieber.
\newblock {Growth of nanotubes for probe microscopy tips}.
\newblock {\em Nature}, 398:761, 1999.

\bibitem{BachtoldSCI2001}
Adrian Bachtold, Peter Hadley, Takeshi Nakanishi, and Cees Dekker.
\newblock {Logic circuits with carbon nanotube transistors}.
\newblock {\em Science}, 294:1317--1320, 2001.

\bibitem{HarneitPSS}
W.~Harneit, C.~Meyer, A.~Weidinger, D.~Suter, and J.~Twamley.
\newblock {}.
\newblock {\em Phys. St. Solidi B}, 233:453, 2002.

\bibitem{HamadaPRL1992}
N.~Hamada, S.~Sawada, and A.~Oshiyama.
\newblock {New one-dimensional conductors: Graphitic microtubules}.
\newblock {\em Phys. Rev. Lett.}, 68:1579.1581, 1992.

\bibitem{DresselhausTubes}
R.~Saito, G.~Dresselhaus, and M.S. Dresselhaus.
\newblock {\em Physical Properties of Carbon Nanotubes}.
\newblock Imperial College Press, 1998.

\bibitem{ChattopadhyayJACS}
D.~Chattopadhyay, L.~Galeska, and F.~Papadimitrakopoulos.
\newblock {A route for bulk separation of semiconducting from metallic
  single-wall carbon nanotubes}.
\newblock {\em J. Am. Chem. Soc.}, 125:3370--3375, 2003.

\bibitem{KrupkeSCI}
R.~Krupke, F.~Hennrich, H.~von Lohneysen, and M.~M. Kappes.
\newblock {Separation of Metallic from Semiconducting Single-Walled Carbon
  Nanotubes}.
\newblock {\em Science}, 301:344--347, 2003.

\bibitem{RinzlerNL}
Z.~H. Chen, X.~Du, M.~H. Du, C.~D. Rancken, H.~P. Cheng, and A.~G. Rinzler.
\newblock {}.
\newblock {\em Nano Lett.}, 3:1245--1259, 2003.

\bibitem{StranoSCI}
M.~Zheng, A.~Jagota, M.~S. Strano, A.~P. Santos, P.~Barone, S.~G. Chou,
  G~Diner, M.~S. B.~A.~Dresselhaus, R.~S. McLean, G.~B. Onoa, G.~G. Samsonidze,
  E.~D. Semke, M.~Usrey, and D.~J. Walls.
\newblock {2003}.
\newblock {\em Science}, 302:1545--1548, 2003.

\bibitem{Bachilo:Science298:2361:(2002)}
Sergei~M. Bachilo, Michael~S. Strano, Carter Kittrell, Robert~H. Hauge,
  Richard~E. Smalley, and R.~Bruce Weisman.
\newblock {Structure-Assigned Optical Spectra of Single-Walled Carbon
  Nanotubes}.
\newblock {\em Science}, 298:2361--2366, 2002.

\bibitem{FantiniPRL2004}
C.~Fantini, A.~Jorio, M.~Souza, M.~S. Strano, M.~S. Dresselhaus, and M.~A.
  Pimenta.
\newblock {Optical Transition Energies for Carbon Nanotubes from Resonant Raman
  Spectroscopy: Environment and Temperature Effects}.
\newblock {\em Phys. Rev. Lett.}, 93:147406, 2004.

\bibitem{TelgPRL2004}
H.~Telg, J.~Maultzsch, S.~Reich, F.~Hennrich, and C.~Thomsen.
\newblock {Chirality Distribution and Transition Energies of Carbon Nanotubes}.
\newblock {\em Phys. Rev. Lett.}, 93:177401, 2004.

\bibitem{EggerPRL1997}
R.~Egger and A.~O. Gogolin.
\newblock {Effective low-energy theory for correlated carbon nanotubes}.
\newblock {\em Nature}, 79:5082–5085, 1997.

\bibitem{BohnenPeierlsPRL2004}
K.~P. Bohnen, R.~Heid, H.~J. Liu, and C.~T. Chan.
\newblock {Lattice Dynamics and Electron-Phonon Interaction in (3,3) Carbon
  Nanotubes}.
\newblock {\em Phys. Rev. Lett.}, 93:245501--1--4, 2004.

\bibitem{ConnetablePeierlsPRL2005}
D.~Conn\'{e}table, G.-M. Rignanese, J.-C. Charlier, and X.~Blase.
\newblock {Room Temperature Peierls Distortion in Small Diameter Nanotubes}.
\newblock {\em Phys. Rev. Lett.}, 94:015503--1--4, 2005.

\bibitem{DekkerNAT1997}
Sander~J. Tans, Michel~H. Devoret, Hongjie Dai, Andreas Thess, Richard~E.
  Smalley, L.~J. Geerligs, and Cees Dekker.
\newblock {Individual single-wall carbon nanotubes as quantum wires}.
\newblock {\em Nature}, 386:474 -- 477, 1997.

\bibitem{KanePRL2003}
C.L. Kane and E.J. Mele.
\newblock {Ratio Problem in Single Carbon Nanotube Fluorescence Spectroscopy}.
\newblock {\em Phys. Rev. Lett.}, 90:207401--1--4, 2003.

\bibitem{LouiePRL2004}
Catalin~D. Spataru, Sohrab Ismail-Beigi, Lorin~X. Benedict, and Steven~G.
  Louie.
\newblock {Excitonic Effects and Optical Spectra of Single-Walled Carbon
  Nanotubes}.
\newblock {\em Phys. Rev. Lett.}, 92:077402--1--4, 2004.

\bibitem{AvourisPRL2004}
Vasili Perebeinos, J.~Tersoff, and Phaedon Avouris.
\newblock {Scaling of Excitons in Carbon Nanotubes}.
\newblock {\em Phys. Phys. Lett.}, 92:257402--1--4, 2004.

\bibitem{AvourisPRL2005}
Vasili Perebeinos, J.~Tersoff, and Phaedon Avouris.
\newblock {Effect of Exciton-Phonon Coupling in the Calculated Optical
  Absorption of Carbon Nanotubes}.
\newblock {\em Phys. Phys. Lett.}, 94:027402--1--4, 2005.

\bibitem{BockrathNAT}
M.~Bockrath, D.~H. Cobden, Jia Lu, Rinzler~A. G., R.~E. Smalley, L.~Balents,
  and P.~L. McEuen.
\newblock {Luttinger-liquid behaviour in carbon nanotubes}.
\newblock {\em Nature}, 397:598 -- 601, 1999.

\bibitem{KatauraNAT2003}
H.~Ishii, H.~Kataura, H.~Shiozawa, H.~Yoshioka, H.~Otsubo, Y.~Takayama,
  T.~Miyahara, S.~Suzuki, Y.~Achiba, M.~Nakatake, T.~Narimura, M.~Higashiguchi,
  K.~Shimada, H.~Namatame, and M.~Taniguchi.
\newblock {Direct observation of Tomonaga-Luttinger-liquid state in carbon
  nanotubes at low temperatures}.
\newblock {\em Nature}, 426:540--544, 2003.

\bibitem{PichlerPRL2004}
H.~Rauf, T.~Pichler, M.~Knupfer, J.~Fink, and H.~Kataura.
\newblock {Transition from a Tomonaga-Luttinger Liquid to a Fermi Liquid in
  Potassium-Intercalated Bundles of Single-Wall Carbon Nanotubes}.
\newblock {\em Phys. Rev. Lett.}, 93:096805--1--4, 2004.

\bibitem{HeinzSCI2005}
Feng Wang, Gordana Dukovic, , Louis~E. Brus, and Tony~F. Heinz.
\newblock {The Optical Resonances in Carbon Nanotubes Arise from Excitons}.
\newblock {\em Science}, 308:838--841, 2005.

\bibitem{MaultzschPRB2005}
J.~Maultzsch, R.~Pomraenke, S.~Reich, E.~Chang, D.~Prezzi, A.~Ruini,
  E.~Molinari, M.~S. Strano, C.~Thomsen, and C.~Lienau.
\newblock {Exciton binding energies in carbon nanotubes from two-photon
  photoluminescence}.
\newblock {\em Phys. Rev. B}, 72:241402, 2005.

\bibitem{SmithNAT}
Brian~W. Smith, Marc Monthioux, and David~E. Luzzi.
\newblock {Encapsulated C$_{60}$ in carbon nanotubes}.
\newblock {\em Nature}, 396:323--324, 1998.

\bibitem{LuzziCPL1999}
B.~W. Smith, M.~Monthioux, and D.E. Luzzi.
\newblock {Carbon nanotube encapsulated fullerenes: a unique class of hybrid
  materials}.
\newblock {\em Chem. Phys. Lett.}, 315:31--36, 1999.

\bibitem{KatauraSM2001}
H.~Kataura, Y.~Maniwa, T.~Kodama, K.~Kikuchi, K.~Hirahara, K.~Suenaga,
  S.~Iijima, S.~Suzuki, Y.~Achiba, and W.~Kr\"atschmer.
\newblock {High-yield fullerene encapsulation in single-wall carbon nanotubes
  }.
\newblock {\em Synthetic Met.}, 121:1195--1196, 2001.

\bibitem{LuzziCPL2000}
B.~W. Smith and D.E. Luzzi.
\newblock {Formation mechanism of fullerene peapods and coaxial tubes: a path
  to large scale synthesis}.
\newblock {\em Chem. Phys. Lett.}, 321:169--174, 2000.

\bibitem{BandowCPL2001}
S.~Bandow, M.~Takizawa, K.~Hirahara, M.~Yudasaka, and S.~Iijima.
\newblock {Raman scattering study of double-wall carbon nanotubes derived from
  the chains of fullerenes in single-wall carbon nanotubes}.
\newblock {\em Chem. Phys. Lett.}, 337:48--54, 2001.

\bibitem{HutchisonCAR2001}
J.~L. Hutchison, N.~A. Kiselev, E.~P. Krinichnaya, A.~V. Krestinin, R.~O.
  Loutfy, A.~P. Morawsky, V.~E. Muradyan, E.~D. Obraztsova, J.~Sloan, S.~V.
  Terekhov, and D.~N. Zakharov.
\newblock {Double-walled carbon nanotubes fabricated by a hydrogen arc
  discharge method}.
\newblock {\em Carbon}, 39:761--770, 2001.

\bibitem{ChengCPL2002}
Wencai Ren, Feng Li, Jian Chen, Shuo Bai, and Hui-Ming Cheng.
\newblock {Morphology, diameter distribution and Raman scattering measurements
  of double-walled carbon nanotubes synthesized by catalytic decomposition of
  methane}.
\newblock {\em Chem. Phys. Lett.}, 359:196--202, 2002.

\bibitem{PfeifferPRL2003}
R.~Pfeiffer, H.~Kuzmany, Ch. Kramberger, Ch. Schaman, T.~Pichler, H.~Kataura,
  Y.~Achiba, J.~K{\"u}rti, and V.~Z{\'o}lyomi.
\newblock {Unusual High Degree of Unperturbed Environment in the Interior of
  Single-Wall Carbon Nanotubes}.
\newblock {\em Phys. Rev. Lett.}, 90:225501--1--4, 2003.

\bibitem{YudasakaCPL}
M.~Yudasaka, K.~Ajima, K.~Suenaga, T.~Ichihashi, A.~Hashimoto, and S.~Iijima.
\newblock { Nano-extraction and nano-condensation for C$_{60}$ incorporation
  into single-wall carbon nanotubes in liquid phases}.
\newblock {\em Chem. Phys. Lett.}, 380:42--46, 2003.

\bibitem{SimonCPL2004}
F.~Simon, H.~Kuzmany, H.~Rauf, T.~Pichler, J.~Bernardi, H.~Peterlik, L.~Korecz,
  F.~F\"ul\"op, and A.~J\'anossy.
\newblock {Low temperature fullerene encapsulation in single wall carbon
  nanotubes: synthesis of N@C$_{60}$@SWCNT}.
\newblock {\em Chem. Phys. Lett.}, 383:362--367, 2004.

\bibitem{Monthioux2004}
M.~Monthioux and L.~No\'{e}.
\newblock {Room temperature synthesis of C$_{60}$@SWNT (peapods)}.
\newblock In {\em {XVIIIth International Winterschool on Electronic Properties
  of Novel Materials.}}, 2004.
\newblock AIP Proceedings.

\bibitem{BriggsJMC}
A.~N. Khlobystov, D.~A. Britz, J.~W. Wang, S.~A. O'Neil, M.~Poliakoff, and
  G.~A.~D. Briggs.
\newblock {Low temperature assembly of fullerene arrays in single-walled carbon
  nanotubes using supercritical fluids}.
\newblock {\em J. Mat. Chem.}, 14:2852--2857, 2004.

\bibitem{PichlerPRL2001}
T.~Pichler, H.~Kuzmany, H.~Kataura, and Y.~Achiba.
\newblock {Metallic Polymers of C$_{60}$ Inside Single-Walled Carbon
  Nanotubes}.
\newblock {\em Phys. Rev. Lett.}, 87:267401--1--4, 2001.

\bibitem{SimonCPL2005}
F.~Simon, \'{A} Kukovecz, Z.~K\'{o}nya, R.~Pfeiffer, and H.~Kuzmany.
\newblock {Highly defect-free inner tubes in CVD prepared double wall carbon
  nanotubes }.
\newblock {\em Chem. Phys. Lett.}, 413:506--511, 2005.

\bibitem{LiuCPL2004}
B.~C. Liu, S.~C. Lyu, S.~I. Jung, H.~K. Kang, C.-W. Yang, J.~W. Park, Park
  C.Y., and C.~J. Lee.
\newblock {Single-walled carbon nanotubes produced by catalytic chemical vapor
  deposition of acetylene over Fe–Mo/MgO catalyst}.
\newblock {\em Chem. Phys. Lett.}, 383:104--108, 2004.

\bibitem{AbePRB2003}
Masatoshi Abe, Hiromichi Kataura, Hiroshi Kira, Takeshi Kodama, Shinzo Suzuki,
  Yohji Achiba, Ken-ichi Kato, Masaki Takata, Akihiko Fujiwara, Kazuyuki
  Matsuda, and Yutaka Maniwa.
\newblock {Structural transformation from single-wall to double-wall carbon
  nanotube bundles}.
\newblock {\em Phys. Rev. B}, 68:041405(R), 2003.

\bibitem{ZolyomiPRB2004}
V.~Z\'{o}lyomi and J.~K\"{u}rti.
\newblock {First-principles calculations for the electronic band structures of
  small diameter single-wall carbon nanotubes}.
\newblock {\em Phys. Rev. B}, 70:085403--1--8, 2004.

\bibitem{EndoNAT}
M.~Endo, H.~Muramatsu, T.~Hayashi, Y.~A. Kim, M.~Terrones, and M.~S.
  Dresselhaus.
\newblock {Nanotechnology: 'buckypaper' from coaxial nanotubes}.
\newblock {\em Nature}, 433:476, 2005.

\bibitem{SimonPRB2005}
F.~Simon, \'A. Kukovecz, C.~Kramberger, R.~Pfeiffer, F.~Hasi, H.~Kuzmany, and
  H.~Kataura.
\newblock {Diameter selective characterization of single-wall carbon
  nanotubes}.
\newblock {\em Phys. Rev. B}, 71:165439--1--5, 2005.

\bibitem{KuzmanyBook}
H.~Kuzmany.
\newblock {\em Solid-State Spectroscopy, An Introduction}.
\newblock Springer Verlag, Berlin, 1998.

\bibitem{SimonPRL2005}
F.~Simon, Ch. Kramberger, R.~Pfeiffer, H.~Kuzmany, J.~Z\'{o}lyomi,
  V.~K\"{u}rti, P.~M. Singer, and H.~Alloul.
\newblock {Isotope Engineering of Carbon Nanotube Systems}.
\newblock {\em Phys. Rev. Lett.}, 95:017401--1--4, 2005.

\bibitem{Kuerti:PhysRevB58:R8869:(1998)}
J.~K\"urti, G.~Kresse, and H.~Kuzmany.
\newblock {First-principles calculations of the radial breathing mode of
  single-wall carbon nanotubes}.
\newblock {\em Phys. Rev. B}, 58:R8869--R8872, 1998.

\bibitem{PfeifferPRB2004}
R.~Pfeiffer, H.~Kuzmany, T.~Pichler, H.~Kataura, Y.~Achiba, M.~Melle-Franco,
  and F.~Zerbetto.
\newblock {Electronic and mechanical coupling between guest and host in carbon
  peapods}.
\newblock {\em Phys. Rev. B}, 69:035404, 2004.

\bibitem{PfeifferPRB2005b}
R.~Pfeiffer, F.~Simon, H.~Kuzmany, and V.~N. Popov.
\newblock {Fine structure of the radial breathing mode of double-wall carbon
  nanotubes}.
\newblock {\em Phys. Rev. B}, 72:161404 --1--4, 2005.

\bibitem{DresselhausTubesNew}
M.~S. Dresselhaus, G.~Dresselhaus, and Ph. Avouris.
\newblock {\em {Carbon Nanotubes: Synthesis, Structure, Properties, and
  Applications}}.
\newblock Springer, Berlin, Heidelberg, New York, 2001.

\bibitem{KatauraSM1999}
H.~Kataura, Y.~Kumazawa, Y.~Maniwa, I.~Umezu, S.~Suzuki, Y.~Ohtsuka, and
  Y.~Achiba.
\newblock {Optical properties of single-wall carbon nanotubes}.
\newblock {\em Synthetic Met.}, 103:2555--2558, 1999.

\bibitem{PfeifferEPJB2004}
R.~Pfeiffer, Ch. Kramberger, F.~Simon, H.~Kuzmany, V.~N. Popov, and H.~Kataura.
\newblock {Interaction between concentric Tubes in DWCNTs}.
\newblock {\em Eur. Phys. J. B}, 42:345--350, 2004.

\bibitem{Popov:PhysRevB65:235415:(2002)}
V.~N. Popov and Luc Henrard.
\newblock {Breathinglike phonon modes of multiwalled carbon nanotubes}.
\newblock {\em Phys. Rev. B}, 65:235415, 2002.

\bibitem{Jorio:PhysRevLett86:1118:(2001)}
A.~Jorio, R.~Saito, J.~H. Hafner, C.~M. Lieber, M.~Hunter, T.~McClure,
  G.~Dresselhaus, and M.~S. Dresselhaus.
\newblock {Structural $(n,m)$ Determination of Isolated Single-Wall Carbon
  Nanotubes by Resonant Raman Scattering}.
\newblock {\em Phys. Rev. Lett.}, 86:1118--1121, 2001.

\bibitem{Thess:Science273:483:(1996)}
Andreas Thess, Roland Lee, Pavel Nikolaev, Hongjie Dai, Pierre Petit, Jerome
  Robert, Chunhui Xu, Young~Hee Lee, Seong~Gon Kim, Andrew~G. Rinzler,
  Daniel~T. Colbert, Gustavo~E. Scuseria, David Tom{\'a}nek, John~E. Fischer,
  and Richard~E. Smalley.
\newblock {Crystalline Ropes of Metallic Carbon Nanotubes}.
\newblock {\em Science}, 273:483--487, 1996.

\bibitem{SimonPRB2006}
F.~Simon, R.~Pfeiffer, and H.~Kuzmany.
\newblock {Temperature dependence of optical excitation life-time and band-gap
  in chirality assigned semiconducting single-wall carbon nanotubes}.
\newblock {\em Phys. Rev. B}, 74:121411(R)--1--4, 2006.

\bibitem{KuzmanyEPJB}
H.~Kuzmany, W.~Plank, M.~Hulman, Ch. Kramberger, A.~Gr\"uneis, Th. Pichler,
  H.~Peterlik, H.~Kataura, and Y.~Achiba.
\newblock {Determination of SWCNT diameters from the Raman response of the
  radial breathing mode}.
\newblock {\em Eur. Phys. J. B}, 22(3):307--320, 2001.

\bibitem{AjayanPRB2002}
Nachiket~R. Raravikar, Pawel Keblinski, Apparao~M. Rao, Mildred~S. Dresselhaus,
  Linda~S. Schadler, and Pulickel~M. Ajayan.
\newblock {Temperature dependence of radial breathing mode Raman frequency of
  single-walled carbon nanotubes}.
\newblock {\em Phys. Rev. B}, 66:235424--1--9, 2002.

\bibitem{KrambergerPRB2003}
Ch. Kramberger, R.~Pfeiffer, H.~Kuzmany, V.~Z\'olyomi, and J.~K\"urti.
\newblock {Assignment of chiral vectors in carbon nanotubes}.
\newblock {\em Phys. Rev. B}, 68:235404, 2003.

\bibitem{SmithCPL2000}
Brian~W. Smith and David~E. Luzzi.
\newblock {Formation mechanism of fullerene peapods and coaxial tubes: a path
  to large scale synthesis}.
\newblock {\em Chem. Phys. Lett.}, 321:169--174, 2000.

\bibitem{LiuPRB2002}
X.~Liu, T.~Pichler, M.~Knupfer, M.~S. Golden, J.~Fink, H.~Kataura, Y.~Achiba,
  K.~Hirahara, and S.~Iijima.
\newblock {Filling factors, structural, and electronic properties of C$_{60}$
  molecules in single-wall carbon nanotubes}.
\newblock {\em Phys. Rev. B}, 65:045419--1--6, 2002.

\bibitem{HasiJNN}
F.~Hasi, F.~Simon, and H.~Kuzmany.
\newblock {Reversible Hole Engineering for Single-Wall Carbon Nanotubes}.
\newblock {\em J. Nanosci. Nanotechn.}, 5:1785--1791, 2005.

\bibitem{SmalleyPRL2002}
Y.~Zhao, B.~I. Yakobson, and R.~E. Smalley.
\newblock {Dynamic Topology of Fullerene Coalescence}.
\newblock {\em Phys. Rev. Lett.}, 88:185501--1--4, 2002.

\bibitem{TomanekPRB}
S.~W. Han, M.~Yoon, S.~Berber, N.~Park, E.~Osawa, J.~Ihm, and D.~Tom\'{a}nek.
\newblock {Microscopic mechanism of fullerene fusion}.
\newblock {\em Phys. Rev. B}, 70:113402--1--4, 2004.

\bibitem{BCS}
J.~Bardeen, L.~N. Cooper, and J.~R. Schrieffer.
\newblock {Theory of Superconductivity}.
\newblock {\em Phys. Rev.}, 108:1175--1204, 1957.

\bibitem{SimonCPL2006}
F.~Simon and H.~Kuzmany.
\newblock {Growth of single-wall carbon nanotubes from $^{13}$C isotope
  labelled organic solvents inside single wall carbon nanotube hosts}.
\newblock {\em Chem. Phys. Lett.}, 425:85--88, 2006.

\bibitem{KressePRB1999}
G.~Kresse and D.~Joubert.
\newblock {From ultrasoft pseudopotentials to the projector augmented-wave
  method}.
\newblock {\em Phys. Rev. B}, 59:1758–1775, 1999.

\bibitem{NemesPRB2000}
A.~S. Claye, N.~M. Nemes, A.~J\'{a}nossy, and J.~E. Fischer.
\newblock {Structure and electronic properties of potassium-doped single-wall
  carbon nanotubes}.
\newblock {\em Phys. Rev. B}, 62:4845--4848 (R), 2000.

\bibitem{SalvetatPRB2005}
J.-P. Salvetat, T.~Feh\'{e}r, C.~L'Huillier, F.~Beuneu, and L.~Forr\'{o}.
\newblock {Anomalous electron spin resonance behavior of single-walled carbon
  nanotubes}.
\newblock {\em Phys. Rev. B}, 72:075440--1--6, 2005.

\bibitem{WeidingerPRL}
T.~Almeida~Murphy, Th. Pawlik, A.~Weidinger, M.~Höhne, R.~Alcala, and J.-M.
  Spaeth.
\newblock {}.
\newblock {\em Phys. Rev. Lett.}, 77:1075, 1996.

\bibitem{WaiblingerPRB}
M.~Waiblinger, K.~Lips, W.~Harneit, A.~Weidinger, E.~Dietel, and A.~Hirsch.
\newblock {Thermal stability of the endohedral fullerenes N@C$_{60}$,
  N@C$_{70}$, and P@C$_{60}$}.
\newblock {\em Phys. Rev. B}, 64:159901--1--4, 2001.

\bibitem{PietzakCPL}
B.~Pietzak, M.~Waiblinger, T.A. Murphy, A.~Weidinger, M.~Hohne, E.~Dietel, and
  A.~Hirsch.
\newblock {}.
\newblock {\em Chem. Phys. Lett.}, 279:259, 1997.

\bibitem{JanossyKirch2000}
A.~J\'{a}nossy, S.~Pekker, F.~F\"{u}l\"{o}p, F.~Simon, and G.~Oszl\'{a}nyi.
\newblock {}.
\newblock In {\em {Proceedings of the IWEPNM}}, page 199, 2000.

\bibitem{TangNMRSCI}
X.-P. Tang, A.~Kleinhammes, H.~Shimoda, L.~Fleming, K.~Y. Bennoune, S.~Sinha,
  C.~Bower, O.~Zhou, and Y.~Wu.
\newblock {Electronic Structures of Single-Walled Carbon Nanotubes Determined
  by NMR}.
\newblock {\em Science}, 288:492, 2000.

\bibitem{DinsePCCP}
K.-P. Dinse.
\newblock {EPR investigation of atoms in chemical traps}.
\newblock {\em Phys. Chem. Chem. Phys.}, 4:5442, 2002.

\bibitem{WudlSCI}
J.~C. Hummelen, B.~Knight, J.~Pavlovich, R.~Gonzalez, and F.~Wudl.
\newblock {Isolation of the Heterofullerene C$_{59}$N as Its Dimer
  (C$_{59}$N)$_{2}$}.
\newblock {\em Science}, 269:1554--1556, 1995.

\bibitem{WudlReview}
J.~C. Hummelen, C.~Bellavia-Lund, and F.~Wudl.
\newblock {\em {Heterofullerenes}}, volume 199, page~93.
\newblock Springer, Berlin, Heidelberg, 1999.

\bibitem{PrassidesPRL}
M.~J. Butcher, F.~H. Jones, P.~H. Beton, P.~Moriarty, B.~N. Cotier, M.~D.
  Upward, K.~Prassides, K.~Kordatos, N.~Tagmatarchis, F.~Wudl, V.~Dhanak, T.~K.
  Johal, C.~Crotti, C.~Comicioli, and C.~Ottaviani.
\newblock {C$_{59}$N Monomers: Stabilization through Immobilization}.
\newblock {\em Phys. Rev. Lett.}, 83:3478--3481, 1999.

\bibitem{FulopCPL}
F.~F\"{u}l\"{o}p, A.~Rockenbauer, F.~Simon, S.~Pekker, L.~Korecz, S.~Garaj, and
  A.~J\'{a}nossy.
\newblock {Azafullerene C$_{59}$N, a Stable Free Radical Substituent in
  Crystalline C$_{60}$}.
\newblock {\em Chem. Phys. Lett.}, 334:223, 2001.

\bibitem{SimonCAR2006}
F.~Simon, H.~Kuzmany, J.~Bernardi, F.~Hauke, and A.~Hirsch.
\newblock {Encapsulating C$_{59}$N azafullerene derivatives inside single-wall
  carbon nanotubes}.
\newblock {\em Carbon}, 44:1958--1962, 2006.

\bibitem{HirschC59NReview}
A.~Hirsch and B.~Nuber.
\newblock {Nitrogen heterofullerenes}.
\newblock {\em Acc. Chem. Res.}, 32:795--804, 1999.

\bibitem{KuzmanyPRB1999}
H.~Kuzmany, W.~Plank, J.~Winter, O.~Dubay, N.~Tagmatarchis, and K.~Prassides.
\newblock {Raman spectrum and stability of (C$_{59}$N)$_{2}$}.
\newblock {\em Phys. Rev. B}, 60:1005--1012, 1999.

\bibitem{SingerPRL2005}
P.~M. Singer, P.~Wzietek, H.~Alloul, F.~Simon, and H.~Kuzmany.
\newblock {NMR Evidence for Gapped Spin Excitations in Metallic Carbon
  Nanotubes }.
\newblock {\em Phys. Rev. Lett.}, 95:236403--1--4, 2005.

\bibitem{SimonPRL2006}
F.~Simon, H.~Kuzmany, B.~N\'{a}fr\'{a}di, T.~Feh\'{e}r, L.~Forr\'{o},
  F.~F\"{u}l\"{o}p, A.~J\'{a}nossy, A.~Rockenbauer, L.~Korecz, F.~Hauke, and
  A.~Hirsch.
\newblock {Magnetic fullerenes inside single-wall carbon nanotubes}.
\newblock {\em Phys. Rev. Lett.}, 97:136801--1--4, 2006.

\bibitem{RockenbauerPRL2005}
A.~Rockenbauer, G.~Cs\'{a}nyi, F.~F\"{u}l\"{o}p, S.~Garaj, L.~Korecz,
  R.~Luk\'{a}cs, F.~Simon, L.~Forr\'{o}, S.~Pekker, and A.~J\'{a}nossy.
\newblock {Electron delocalization and dimerization in solid C$_{59}$N doped
  C$_{60}$ fullerene}.
\newblock {\em Phys. Rev. Lett.}, 94:066603, 2005.

\bibitem{SimonJCP}
F.~Simon, D.~Ar\v{c}on, N.~Tagmatarchis, S.~Garaj, L.~Forr\'{o}, and
  K.~Prassides.
\newblock {ESR signal in azafullerene (C$_{59}$N)$_{2}$ induced by thermal
  homolysis}.
\newblock {\em J. Phys. Chem. A.}, 103:6969, 1999.

\bibitem{GozeBacCAR2002}
C.~Goze-Bac, S.~Latil, P.~Lauginie, V.~Jourdain, J.~Conard, L.~Duclaux,
  A.~Rubio, and P.~Bernier.
\newblock {Magnetic interactions in carbon nanostructures}.
\newblock {\em Carbon}, 40:1825--1842, 2002.

\bibitem{SimonDWCNTReview}
F.~Simon, R.~Pfeiffer, C.~Kramberger, M.~Holz\-weber, and H.~Kuzmany.
\newblock {\em {The Raman response of double wall carbon nanotubes}}, pages
  203--224.
\newblock Springer New York, 2005.

\bibitem{KuertiNJP2003}
J.~K{\"u}rti, V.~Z{\'o}lyomi, M.~Kert{\'e}sz, and S.~Guangyu.
\newblock {The geometry and the radial breathing mode of carbon nanotubes:
  beyond the ideal behaviour}.
\newblock {\em New. J. Phys.}, 5:125.1–125.21, 2003.

\bibitem{MauriPRB2006}
M.~A.~L. Marques, M.~d'Avezac, and F.~Mauri.
\newblock {Magnetic response of carbon nanotubes from ab initio calculations},
  2006.

\bibitem{SlichterBook}
C.~P. Slichter.
\newblock {\em Principles of Magnetic Resonance}.
\newblock Spinger-Verlag, New York, 3rd ed. 1996 edition, 1989.

\bibitem{ShimodaPRL2002}
H.~Shimoda, B.~Gao, X.~P. Tang, A.~Kleinhammes, L.~Fleming, Y.~Wu, and O.~Zhou.
\newblock {Lithium Intercalation into Opened Single-Wall Carbon Nanotubes:
  Storage Capacity and Electronic Properties}.
\newblock {\em Phys. Rev. Lett.}, 88:15502, 2002.

\bibitem{KleinhammesPRB2003}
A.~Kleinhammes, S.-H. Mao, X.-J. Yang, X.-P. Tang, H.~Shimoda, J.~P. Lu,
  O.~Zhou, and Y.~Wu.
\newblock {Gas adsorption in single-walled carbon nanotubes studied by NMR}.
\newblock {\em Phys. Rev. B}, 68:75418--1--6, 2003.

\bibitem{OkadaPRL2003}
S.~Okada and A.~Oshiyama.
\newblock {Curvature-Induced Metallization of Double-Walled Semiconducting
  Zigzag Carbon Nanotubes}.
\newblock {\em Phys. Rev. Lett.}, 91:216801--1--4, 2003.

\bibitem{Moriya}
T.~Moriya.
\newblock {The Effect of Electron-Electron Interaction on the Nuclear Spin
  Relaxation in Metals}.
\newblock {\em J. Phys. Soc. Jpn.}, 18:516, 1963.

\bibitem{LuPRL1995}
Jian~Ping Lu.
\newblock {Novel Magnetic Properties of Carbon Nanotubes}.
\newblock {\em Phys. Rev. B}, 74:1123, 1995.

\bibitem{KanePRL1997}
C.L. Kane and E.J. Mele.
\newblock {Size, Shape, and Low Energy Electronic Structure of Carbon
  Nanotubes}.
\newblock {\em Phys. Rev. Lett.}, 78:1932--1935, 1997.

\bibitem{MintmirePRL1998}
J.W. Mintmire and C.T. White.
\newblock {Universal Density of States for Carbon Nanotubes}.
\newblock {\em Phys. Rev. Lett.}, 81:2506, 1998.

\bibitem{Yoshioka}
H.~Yoshioka.
\newblock {NMR relaxation rate in metallic carbon nanotubes}.
\newblock {\em J. Phys. Chem. Solids.}, 63:1281, 2002.

\end{thebibliography}

\end{document}